\documentclass[preprint,aps,preprintnumbers,amsmath,amssymb,superscriptaddress]{revtex4}
\usepackage[version=4]{mhchem}
\usepackage{graphicx}
\usepackage{hyperref}
\usepackage{color}
\usepackage{ulem}
\usepackage{times}
\usepackage{soul}
\usepackage{pdfpages}
\begin{document}

\title{Relating spin-polarized STM imaging and inelastic neutron scattering in the van-der-Waals ferromagnet \ce{Fe3GeTe2}}
\author{Christopher Trainer}
\affiliation{SUPA, School of Physics and Astronomy, University of St Andrews, North Haugh, St Andrews, Fife, KY16 9SS, United Kingdom}
\author{Olivia R. Armitage}
\affiliation{SUPA, School of Physics and Astronomy, University of St Andrews, North Haugh, St Andrews, Fife, KY16 9SS, United Kingdom}
\author{Harry Lane}
\affiliation{School of Physics and Astronomy, University of Edinburgh, Edinburgh EH9 3JZ, United Kingdom}
\affiliation{School of Chemistry and Centre for Science at Extreme Conditions, University of Edinburgh, Edinburgh EH9 3FJ, United Kingdom}
\affiliation{ISIS Pulsed Neutron and Muon Source, STFC Rutherford Appleton Laboratory, Harwell Campus, Didcot, Oxon, OX11 0QX, United Kingdom}
\author{Luke C. Rhodes}
\affiliation{SUPA, School of Physics and Astronomy, University of St Andrews, North Haugh, St Andrews, Fife, KY16 9SS, United Kingdom}
\author{Edmond Chan}
\affiliation{School of Physics and Astronomy, University of Edinburgh, Edinburgh EH9 3JZ, United Kingdom}
\affiliation{Institute Laue-Langevin, 6 rue Jules Horowitz, Boite Postale 156, 38042 Grenoble Cedex 9, France}
\author{Izidor Benedi\v{c}i\v{c}}
\affiliation{SUPA, School of Physics and Astronomy, University of St Andrews, North Haugh, St Andrews, Fife, KY16 9SS, United Kingdom}
\author{J. A. Rodriguez-Rivera}
\affiliation{NIST Center for Neutron Research, National Institute of Standards and Technology, Gaithersburg, Maryland 20899, USA}
\affiliation{Department of Materials Science and Engineering, University of Maryland, College Park, Maryland 20742, USA}
\author{O. Fabelo}
\affiliation{Institute Laue-Langevin, 6 rue Jules Horowitz, Boite Postale 156, 38042 Grenoble Cedex 9, France}
\author{Chris Stock}
\affiliation{School of Physics and Astronomy, University of Edinburgh, Edinburgh EH9 3JZ, United Kingdom}
\author{Peter Wahl}
\affiliation{SUPA, School of Physics and Astronomy, University of St Andrews, North Haugh, St Andrews, Fife, KY16 9SS, United Kingdom}

\date{\today}

\begin{abstract} 
Van-der-Waals (vdW) ferromagnets have enabled the development of heterostructures assembled from exfoliated monolayers with spintronics functionalities, making it important to understand and ultimately tune their magnetic properties at the microscopic level. Information about the magnetic properties of these systems comes so far largely from macroscopic techniques, with little being known about the microscopic magnetic properties. Here, we combine spin-polarized scanning tunneling microscopy and quasi-particle interference imaging with neutron scattering to establish the magnetic and electronic properties of the metallic vdW ferromagnet $\mathrm{Fe_3GeTe_2}$. By imaging domain walls at the atomic scale, we can relate the domain wall width to the exchange interaction and magnetic anisotropy extracted from the magnon dispersion as measured in inelastic neutron scattering, with excellent agreement between the two techniques. From comparison with Density Functional Theory calculations we can assign the quasi-particle interference to be dominated by spin-majority bands. We find a dimensional dichotomy of the bands at the Fermi energy: bands of minority character are predominantly two-dimensional in character, whereas the bands of majority character are three-dimensional. We expect that this will enable new design principles for spintronics devices.
\end{abstract}
\maketitle

\section{Introduction}

The discovery of ferromagnetic van-der-Waals (vdW) materials has enabled the possibility of manufacturing spintronics devices from vdW heterostructures \cite{He_2021}. Although, according to the Mermin-Wagner theorem, ferromagnetism should be unstable in two dimensions, recently a number of materials, including \ce{Fe3GeTe2}, have been shown to exhibit ferromagnetism down to the monolayer limit \cite{Gong2017,Huang2017,Fei_2021}. The evolution of its magnetic order from 3D to 2D is an interesting open question \cite{Deng2018}. The study of magnetic properties in 2D materials and at surfaces, however, is challenging. Conventional methods used to establish magnetic order parameters, such as neutron scattering, are not suitable for monolayer-thin samples and surfaces. Here, we use spin-polarised STM, quasiparticle interference imaging and neutron scattering to elucidate the interplay between the bulk and surface magnetic order and the low energy electronic structure of a quasi-2D ferromagnet. We identify two types of defects arising predominantly from Fe and Te vacancies, and show that quasiparticle scattering from these defects produces magnetic scattering dominated by the more two-dimensional electronic bands around the Fermi level. Imaging of a domain wall and comparison of its profile with the exchange coupling $J$ and magnetic anisotropy $K$ obtained from inelastic neutron scattering of bulk \ce{Fe_3GeTe_2} reveals good agreement, with evidence for a larger magnetic anisotropy and smaller exchange coupling in the surface layer.


\begin{figure}[bt!]
\begin{center}
\includegraphics{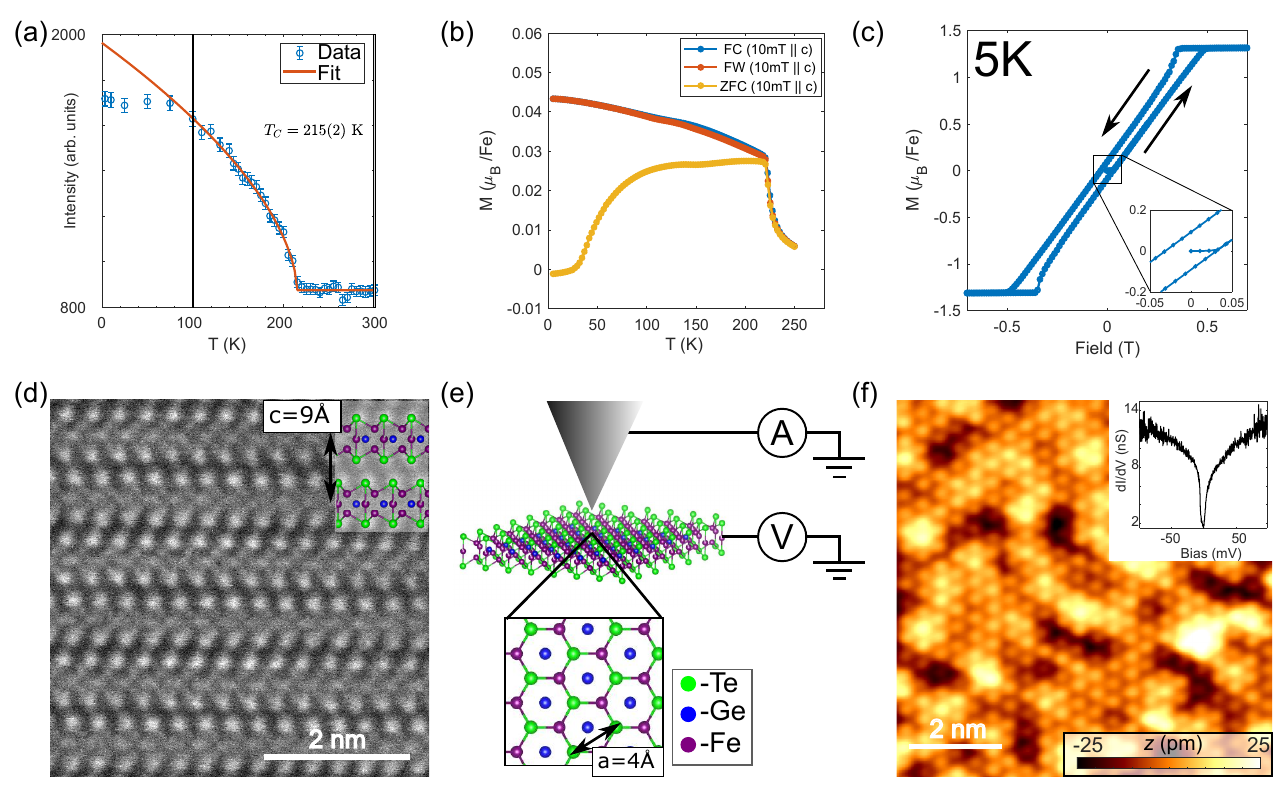}
\end{center}
\caption{\textbf{Sample characterisation.} (a) Intensity of neutrons scattered at the atomic Bragg peak as a function of temperature.  The signal will consist of the structural peak and the magnetic signal due to the ferromagnetic order. The red line represents a power law fit to the data giving a ferromagnetic transition temperature of $215\mathrm{K}$. (b) Magnetization $M$ of the \ce{Fe3GeTe2} crystal used for the STM measurements. Zero field cooled (ZFC), field cooled (FC) and field warmed (FW) measurements are shown. The transition temperature $T_\mathrm{c}$ is found to be $T_\mathrm{c}=218\mathrm{K}$ and is extracted by fitting a Curie Weiss law to the high temperature data. (c) A low temperature magnetization $M$ vs. field $H$ loop recorded in the ferromagnetic phase. (d) TEM image of the layered structure of \ce{Fe3GeTe2}. Inset - side view of the crystal structure. (e) Schematic of the STM tunnel junction experimental set up. Inset the \ce{Fe3GeTe2} crystal structure. (f) A topographic STM image of the \ce{Fe3GeTe2} surface. Inset - Typical STS spectrum recorded on the surface of \ce{Fe3GeTe2}.    
}
\label{topography}
\end{figure}

\section{Results}
\subsection{Sample characterization}
\ce{Fe3GeTe2} has a layered crystal structure with weak interlayer interactions and becomes ferromagnetic below $T=230\mathrm K $\cite{deiseroth_fe3gete2_2006}. The material typically comes with an off-stoichiometry, where the iron concentration deviates from three due to vacancies and interstitial iron. The magnetic properties vary with the excess iron concentration \cite{may_magnetic_2016}. Previous STM and ARPES investigations have been interpreted as Kondo-lattice-like behaviour \cite{zhang_emergence_2018}. The material cleaves easily between the \ce{Fe_{3-y}GeTe2} layers, exhibiting a Te-terminated surface. We have used single crystal neutron diffraction to determine the crystal structure and in particular the exact stoichiometry of our samples, revealing an iron deficiency $y=0.14$ (see suppl. section I).

Magnetic characterisation (Fig.~\ref{topography}(a, b)) reveals a behaviour in field-cooled measurements consistent between the neutron scattering intensity of the Bragg peak and magnetisation, as well as with previous reports \cite{Tian2019}. We find the magnetic transition at about $220\mathrm K$ and a suppression of magnetisation for zero-field cooled measurements starting below $100\mathrm K$, Fig.~\ref{topography}(b), indicating stabilization of a domain structure which results in zero net magnetisation. In measurements of the magnetization $M$ as a function of field $H$, shown in Fig.~\ref{topography}(c), we find magnetic hysteresis, as expected for a ferromagnet. 

Cross-sectional TEM images (Fig.~\ref{topography}(d)) show the stacking sequence as expected from the crystal structure and the high quality of the samples, confirming the AB layer stacking. This type of stacking results in a Rashba-like band crossing at the K point and leads to the formation of topological line nodes \cite{Kim2018}.

Fig.~\ref{topography}(e) shows the measurement set up for the STM measurement and expected surface termination. Topographic imaging of the \ce{Fe3GeTe2} surface (Fig.~\ref{topography}(f)) reveals a hexagonal lattice which can be attributed to the uppermost \ce{Te} lattice. The surface further exhibits a pronounced electronic inhomogeneity, which likely originates from the off-stoichiometry of the sample due to the iron deficiency. STS spectra show a pronounced gap around the Fermi level (inset of Fig.~\ref{topography}(f)). 

\begin{figure}[bt!]
\begin{center}
\includegraphics[scale=0.5]{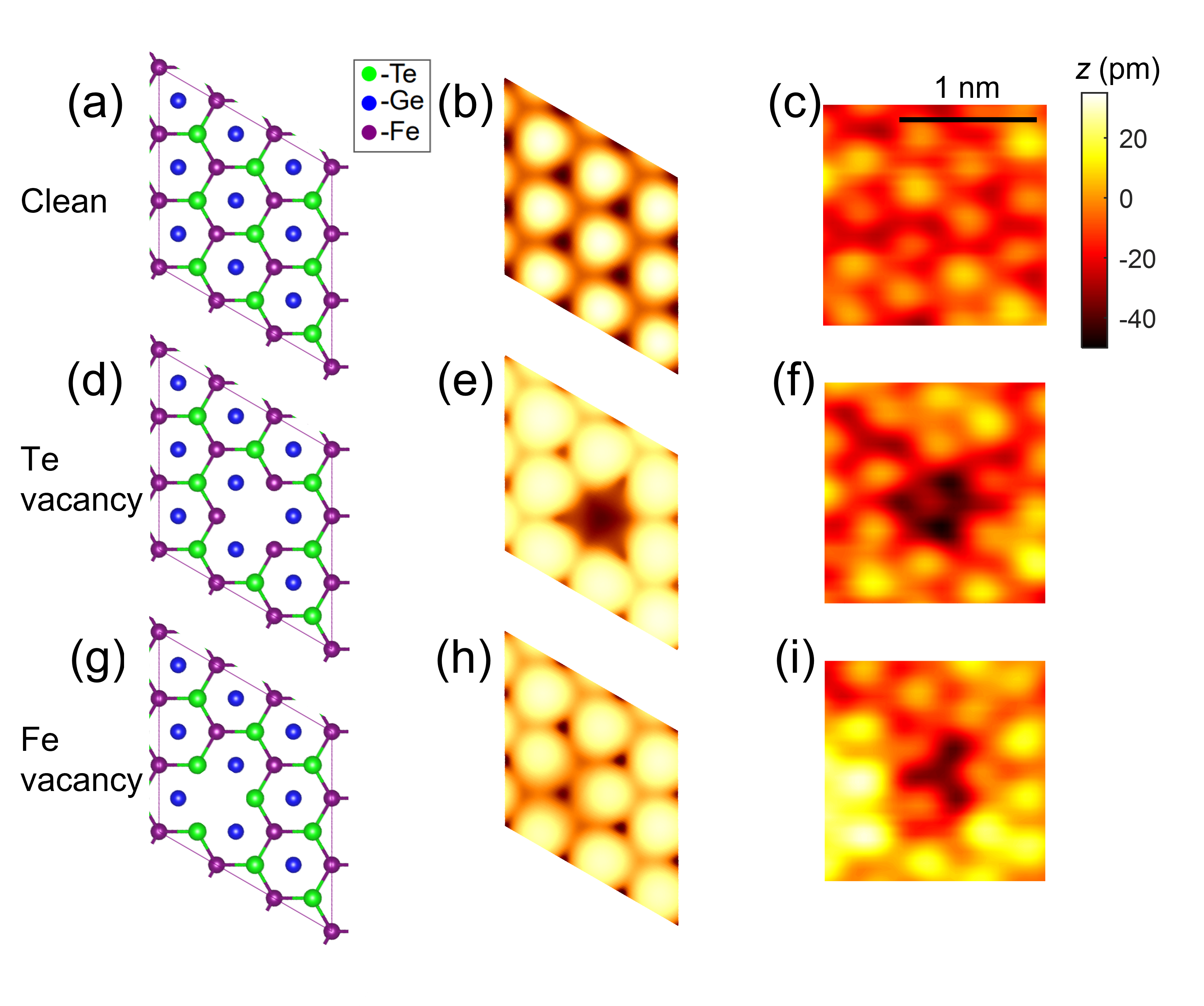}
\end{center}
\caption{\textbf{Identification of defects.} (a) Top view of a $3 \times 3$ supercell of a monolayer of \ce{Fe3GeTe2}. (b) Simulated STM topography of the supercell in (a). (c) STM topography of a region of the \ce{Fe3GeTe2} surface containing no defects. (d) Top view of supercell in (a) with a Te atom removed. (e) Simulated STM topography of the supercell in (d). (f) STM topography of the \ce{Fe3GeTe2} surface, showing a Te-site defect. (g) Same supercell as in (a), with an Fe atom removed, corresponding to $y=0.14$ (i.e. \ce{Fe_{2.86}GeTe2}). (h) Simulated STM topography of the supercell in (g). (i) STM topography of the  \ce{Fe3GeTe2} surface, showing an Fe-site defect. STM images (c), (f) and (i) taken with $V_s=50\mathrm{mV}$, $I=50\mathrm{pA}$. Simulated images (b), (e) and (h) calculated for constant current and $V_s=100\mathrm{mV}$.}
\label{defecttopos}
\end{figure}

To gain a better understanding of the inhomogeneity in our topographic images, we have simulated STM images for different types of defects. Fig.~\ref{defecttopos} shows DFT calculations and measured topographic images for the clean surface and two types of defects, an iron vacancy and a tellurium vacancy. Comparison of the DFT calculations with the STM image for the clean surface (Fig.~\ref{defecttopos}(a)-(c)) suggests that the atomic corrugation in the STM images is due to the surface tellurium lattice. A vacancy of a tellurium atom (as indicated in Fig. \ref{defecttopos}(d)) results in a clear triangular-like defect (Fig.~\ref{defecttopos}(e)) which can be easily identified in the topographic image (Fig.~\ref{defecttopos}(f)). 
An Fe-vacancy, shown in Fig.~\ref{defecttopos}(g), produces a more subtle depression of the local density of states, as shown in Fig.~\ref{defecttopos}(h), which can nevertheless be easily identified in our STM topographies (Fig.~\ref{defecttopos}(i)). This depression in the density of states is likely responsible for the inhomogeneity in topographic images as shown in Fig.~\ref{topography}(b), and has been shown to be responsible for the lowering in $T_c$ as a function of Fe deficiency\cite{may_magnetic_2016}. Interestingly, within our ferromagnetic DFT calculations for a monolayer of \ce{Fe3GeTe2} in a $3\times 3$ supercell, we find that the total magnetic moment per Fe atom is reduced by the presence of a defect (2.004 $\mu_\mathrm B$/Fe vs. 2.123 $\mu_\mathrm B$/Fe without a defect, see the method section for details). 

\subsection{Magnetic properties from STM and neutron scattering}
We have studied the magnetism in \ce{Fe3GeTe2} using neutron scattering and spin-polarized STM to establish a better understanding of its magnetic properties, and the effect of the vacuum interface which will become more important in the 2D limit.  

\begin{figure}[bt!]
\begin{center}
\includegraphics{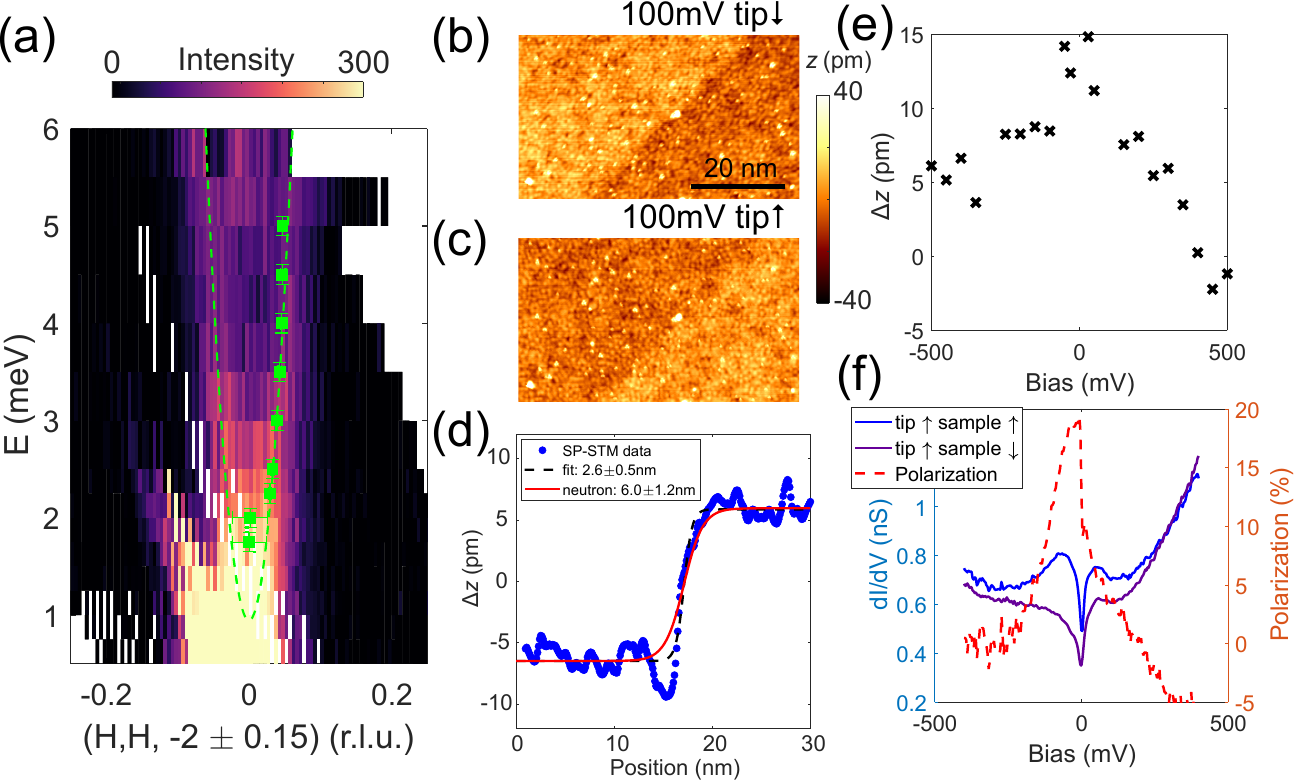}
\end{center}
\caption{\textbf{Magnon dispersion and magnetic imaging.} (a) Inelastic neutron scattering measurement of the spin-wave dispersion of \ce{Fe3GeTe2} around $\mathbf{q}=(H,H,-2)$ for $H=-0.2\ldots 0.2$. The green line represents a parabolic fit to the data. The fit results in $J\approx 43(10)\mathrm{meV}$ and $K\approx 0.6(1)\mathrm{meV}$ (see main text for details).  (b) A ferromagnetic domain wall imaged using spin polarized STM ($V_s=100\mathrm{mV}$, $I_s=125\mathrm{pA}$). The domain wall runs diagonally through the image.  (c) The same area as in (b) imaged with a spin-polarized tip with the opposite spin polarization from that used in (b) ($V_s=100\mathrm{mV}$, $I=125\mathrm{pA}$). (d) A line profile $z(x)$ taken through the difference of images (b) and (c) perpendicular to the domain wall. The red line shows the expected profile from the neutron scattering measurement. (e) The height difference $\Delta z$ recorded between oppositely polarized areas as a function of applied bias voltage. (f) dI/dV spectra (blue curve) recorded on either side of the domain wall shown in (b) and (c) ($V_s=400\mathrm{mV}$, $I=250\mathrm{pA}$, $V_{\mathrm{mod}}=3\mathrm{mV}$). The spectroscopy setpoint was chosen at a bias voltage where the domain wall was not visible. The resulting spin polarization determined from the dI/dV spectra is also shown (red curve).}
\label{magneticcontrast}
\end{figure}

Fig.~\ref{magneticcontrast}(a) shows the inelastic neutron data obtained from a single crystal, with a clear magnon dispersion in a cut through the $(0,0,2)$ peak starting from about $1\mathrm{meV}$. At long wavelengths, the magnetic excitations of an itinerant ferromagnet are well-described by a model of localized magnetic moments \cite{Lovesey}. To describe the spin wave dispersion in \ce{Fe3GeTe2} at small $\mathbf{q}$, we use an effective two-dimensional model of interacting Fe$^{3+}$ ions on a honeycomb lattice with an easy-axis anisotropy, \begin{equation}
\mathcal{H}=J\sum_{\mathbf{r}}^{j \in\{0,1,2\}}\mathbf{S}_{A}(\mathbf{r})\cdot\mathbf{S}_{B}(\mathbf{r}+\mathbf{a}_{j})+K[(\hat{S}_{A}^{z}(\mathbf{r}))^{2}+(\hat{S}_{B}^{z}(\mathbf{r}))^{2}].
\end{equation}
Here, $J$ is the nearest neighbour interaction, $K$ the magnetic anisotropy, $\mathbf{S}_{A,B}$ is the spin operator acting on site $A$ or $B$ and the vectors $\mathbf{a}_{0}=(0,0)$, $\mathbf{a}_{1}=(0,1)$ and $\mathbf{a}_{2}=(1,1)$ span between the unit cells containing nearest neighbor spins (in the $P6_{3}/mmc$ space group). We have neglected further nearest neighbors since their effect in the limit of small $|\mathbf{q}|$ is a simple renormalization of the spin-wave velocity and anisotropy gap. The eigenvalues obtained from the model are $E^{\pm}_{\mathbf{q}}=v_{S}\pm |\gamma_{\mathbf{q}}|$, with $v_{S}=-(3JS+2KS)$ and $\gamma_{\mathbf{q}}=JS\sum_{j}e^{i\mathbf{q}\cdot \mathbf{a}_{j}}$, where $S$ is the spin \cite{Owerre,Pershoguba}. With a wavevector of $\mathbf{q} = (H,H)$, where $H$ is here used to parametrize the reciprocal lattice vector, the dispersion is quadratic for small $H$, $E\approx v_{S}\pm3JS\pm 4\pi^{2} JS H^{2}$. We have fit Gaussian peaks to a series of constant energy cuts in the $(H,H,-2)$ plane through the MACS data to obtain the magnon dispersion. The exchange coupling can be determined from the prefactor, $4\pi^{2}JS$, of the quadratic term. Taking the reduced value of $S=0.8(1)$, the fit yields $J\approx43(10)$ meV and $K\approx0.6(1)$ meV. The magnetic anisotropy is non-negligible, consistent with previous reports \cite{Deng2018,Tan2018}, though somewhat smaller than the one reported previously from neutron scattering, whereas the exchange coupling is larger \cite{exchangecouple}. The difference in these values suggests a strong influence of the iron deficiency on the magnetic excitations: while our crystal has $y=0.14$, the one studied in Ref.~\onlinecite{exchangecouple} had $y=0.25$, almost twice that of our sample.


We can compare these values with the magnetic properties of the surface layer. Using spin-polarized STM\cite{Wiesendangerrev,Bode_2003} we were able to directly image ferromagnetic domain walls at the surface of \ce{Fe3GeTe2} (Fig.~\ref{magneticcontrast}(a)). 
The tip cluster was found to have a sufficiently small moment so that the tip magnetisation can be switched by the magnetic interaction with the sample. 
Manipulating the tip magnetisation in this way allowed us to image the same area with an oppositely polarized tip without having to apply an external magnetic field which would at the same time move the domain wall and polarize the sample. Fig.~\ref{magneticcontrast}(c) shows an image with the same tip after the tip magnetisation has flipped. By taking the difference of the images, we obtain an image of only the magnetic contrast (Fig.~\ref{magneticcontrast}(d)). From a line profile normal to the domain wall (Fig.~ \ref{magneticcontrast}(d)), we can analyze its width by fitting $z(x)=a+b\mathrm{tanh}(\pi(x-x_0)/\delta)$ to the data \cite{hexagonaldelta,aharoni,domain1,domain2,Wiesendangerrev}, assuming that the tip magnetization is parallel to the out-of-plane direction and reverses by $180^\circ$ when changing its direction. We obtain a domain wall width of $\delta=2.6\pm 0.5\mathrm{nm}$. The domain wall width found here is smaller than that typically found in ferromagnetic films which are usually on the order of $10\sim100\mathrm{nm}$ \cite{domain1,domain2}. We can compare the domain wall width $\delta$ with the expected width $\delta_n$ using the exchange interaction $J$ and magnetic anisotropy $K$ obtained from neutron scattering through $\delta_n=\pi\sqrt{\frac{A}{K}}$ \cite{Blundell}, where $A$ is the exchange stiffness of the spins of the system being considered which is related to the exchange interaction $J$. For a 2D hexagonal honeycomb lattice, this expression becomes $\delta_n=\pi d_{nn}\sqrt{\frac{3J}{4K}}$ \cite{hexagonaldelta,aharoni} (see supplementary section S4), where $d_{nn}$ is the nearest neighbour distance between \ce{Fe} atoms. By taking the values of $J$ and $K$ determined from the magnon dispersion in Fig.~\ref{magneticcontrast}(a), we find a domain wall width $\delta_n= 6.0\pm1.2\mathrm{nm}$ which is somewhat larger than what is obtained from a fit to our spin-polarized STM measurement (see fig.~\ref{magneticcontrast}(d)). We note, however, that visually, the domain wall profile suggested by neutron scattering appears very similar to the observed one. The smaller domain wall width which we observe at the surface suggests that the exchange coupling $J$ is smaller in the surface layer compared to the bulk, and the magnetic anisotropy $K$ larger.\\ 
The apparent height of the domain wall in the spin polarized STM images exhibits a strong dependence on the bias voltage. From images recorded at different bias voltages, we can determine how the spin polarization of the sample's electronic structure evolves with energy. For images recorded with a low bias voltage ($<50\mathrm{mV}$) the domain wall shows the largest contrast. As the bias voltage is increased the magnetic contrast decreases and disappears at $400\mathrm{mV}$ before reversing at even higher bias values. For negative applied bias the magnetic contrast remains more or less constant. By recording STS spectra on either side of the domain wall (hence with different magnetization directions of the sample) with a set point condition where the domain wall is not visible it is possible to extract the spin polarization as a function of bias voltage. The  spectra recorded on either side of the domain wall are shown in Fig.~\ref{magneticcontrast}(f).  The polarization can be extracted from the spectra using the relation $P=\frac{g_{\uparrow\uparrow}(V)-g_{\uparrow\downarrow}(V)}{g_{\uparrow\uparrow}(V)+g_{\uparrow\downarrow}(V)}$\cite{Bode_2003,Wiesendangerrev} (Fig.~\ref{magneticcontrast}(f)). The spin polarization shows a sharp peak  of up to $20\%$ spin polarization just below the Fermi level at an energy of $-43\pm6\mathrm{mV}$.        

\begin{figure}[bt!]
\begin{center}
\includegraphics[scale=0.45]{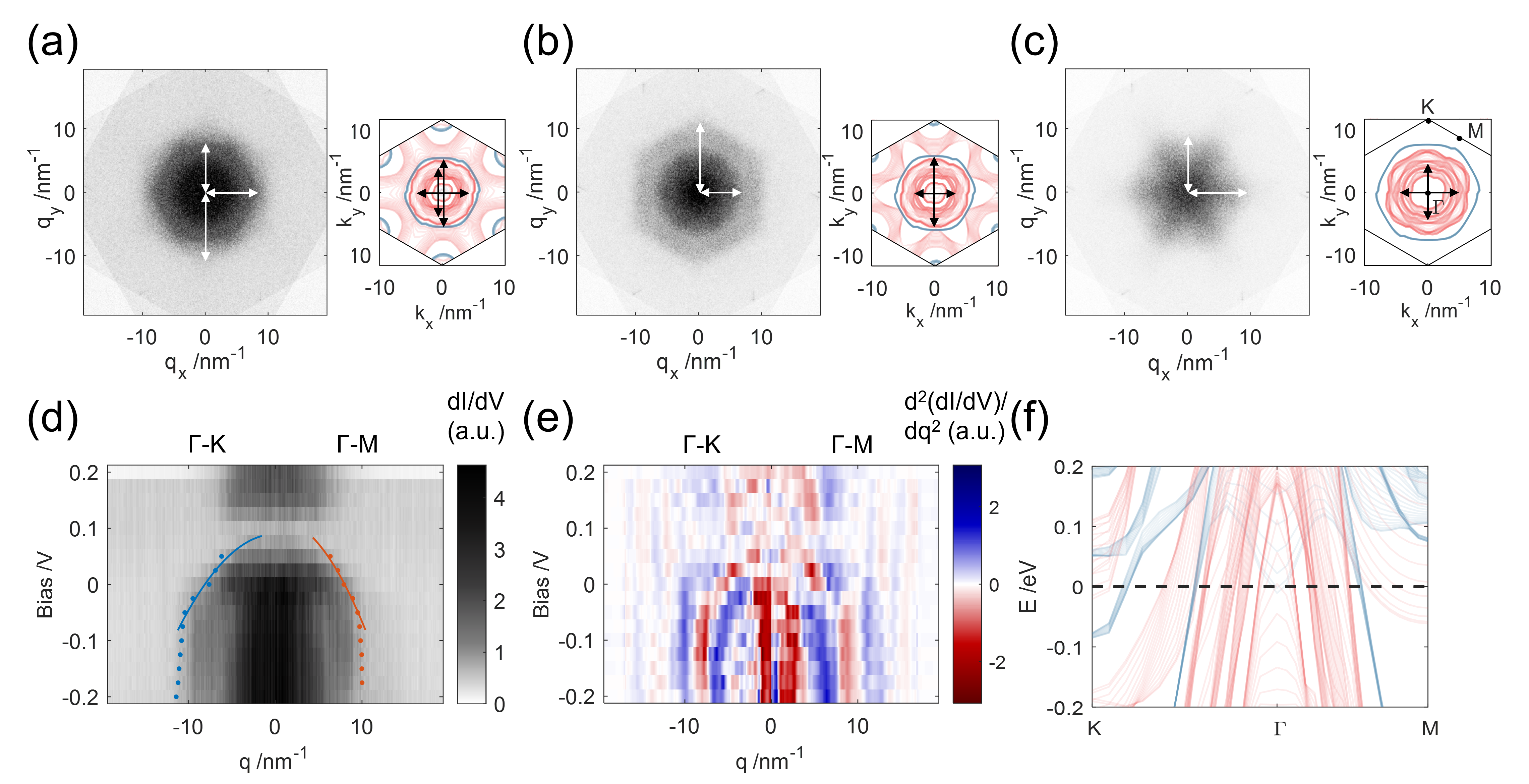}
\end{center}
\caption{\textbf{Quasiparticle interference imaging.} (a-c) Six-fold symmetrised FFT of the differential conductance $\tilde{g}(\mathbf{q},V)$ and calculated constant energy contours, showing majority (red) and minority (blue) spin bands averaged over 16 $k_z$ values from 0 to 2$\pi/c$, at (a) $0\mathrm{mV}$, (b) $-60\mathrm{mV}$ and (c) $-300\mathrm{mV}$. Corresponding scattering vectors are marked on the experimental data and band structure calculation. (d) $\tilde{g}(\mathbf{q},V)$ as a function of bias $V$ for $\mathbf{q}$ along $\Gamma$-K and $\Gamma$-M taken from  similar $\tilde{g}(\mathbf{q},V)$ measurements to (a)-(c), averaged between $q_x/q_y = \pm0.378 \mathrm{nm}^{-1}$ for $\Gamma$-K/$\Gamma$-M. Points fitted to the band dispersions along both directions are also plotted, with parabolic fits to the highest four energy points for each direction giving an average effective mass of $m^\ast = 5.3\pm0.8m_e$ ($\Gamma$-M: $4.5m_e$, $\Gamma$-K: $6.1m_e$). (e) Second derivative with respect to $\mathbf{q}$ of the differential conductance map, $\frac{\partial^{2}\tilde{g}(\mathbf{q},V)}{\partial q^{2}}$, as a function of bias for $\mathbf{q}$ along $\Gamma$-K and $\Gamma$-M. The $\tilde{g}(\mathbf{q},V)$ data was smoothed along the $\mathbf{q}$-direction with a window of $5.060\mathrm{nm}^{-1}$ before calculating the second derivative. (f) DFT calculation of the band structure of \ce{Fe3GeTe2}, showing majority (red) and minority (blue) spin bands averaged over 16 $k_z$ values from 0 to $2\pi/c$.}
\label{qpi}
\end{figure}

\subsection{Quasi-particle interference}
To characterize the interplay between magnetism and the electronic structure, we have used quasi-particle interference imaging. Fig.~\ref{qpi}(a)-(c) show the Fourier transform of quasi-particle interference maps $\tilde{g}(\mathbf{q},V)$ at three different bias voltages ($0\mathrm{mV}$, $-60\mathrm{mV}$ and $-300\mathrm{mV}$), with the calculated band structure in the first Brillouin zone for comparison, integrated over $k_z$. There are clear hexagonal features in the experimental data, marked with arrows, with equivalent scattering vectors shown in the calculated band structure. All of the vectors observed experimentally can be assigned to scattering processes between bands with majority spin character. These bands are also the ones which exhibit a more two-dimensional character, as represented by the opacity in the calculation. Despite being strongly two-dimensional, the minority band centred around the $\Gamma$ point is not visible in any of our measurements. The DFT calculations show that the spin-minority bands have predominantly iron character with little weight on the outer-most Te atoms, whereas the majority bands have Te character, suggesting that we fail to detect the minority band due to matrix element effects. 

A cut of the differential conductance along K-$\Gamma$-M in the 2D Brillouin zone is shown in Fig.~\ref{qpi}(d) between $-200\mathrm{mV}$ and $+200\mathrm{mV}$. From the QPI cuts we can extract the properties of the hole-like band that crosses the Fermi level. By fitting a parabolic dispersion, Fig.~\ref{qpi}(d), and assuming that it is due to intra-band scattering we determine that it has an effective mass $m^\ast = 5.3\pm0.8m_e$ and a Fermi wavevector $k_f=4.1\pm0.3\mathrm{nm}^{-1}$ along $\Gamma - \mathrm{K}$ ($8.2\mathrm{nm}^{-1}$ in q-space) which is in approximate agreement with that of a hexagonal pocket centered around $\Gamma$ seen in ARPES measurements \cite{Kim2018,zhang_emergence_2018,PRBARPES}. From the fits we find a band maximum at $106\pm25\mathrm{mV}$ above the Fermi level. To highlight sharp features, the second derivative with respect to $\mathbf{q}$ of the smoothed dI/dV data is plotted in Fig.~\ref{qpi}(e). In these, the hole band is more readily observed with a maximum at around $150\mathrm{mV}$ at the $\Gamma$ point. We assign this band to the two-dimensional band at the same energy in the calculated $k_z$-integrated band structure, Fig.~\ref{qpi}(f). 

\begin{figure}
    \centering
    \includegraphics{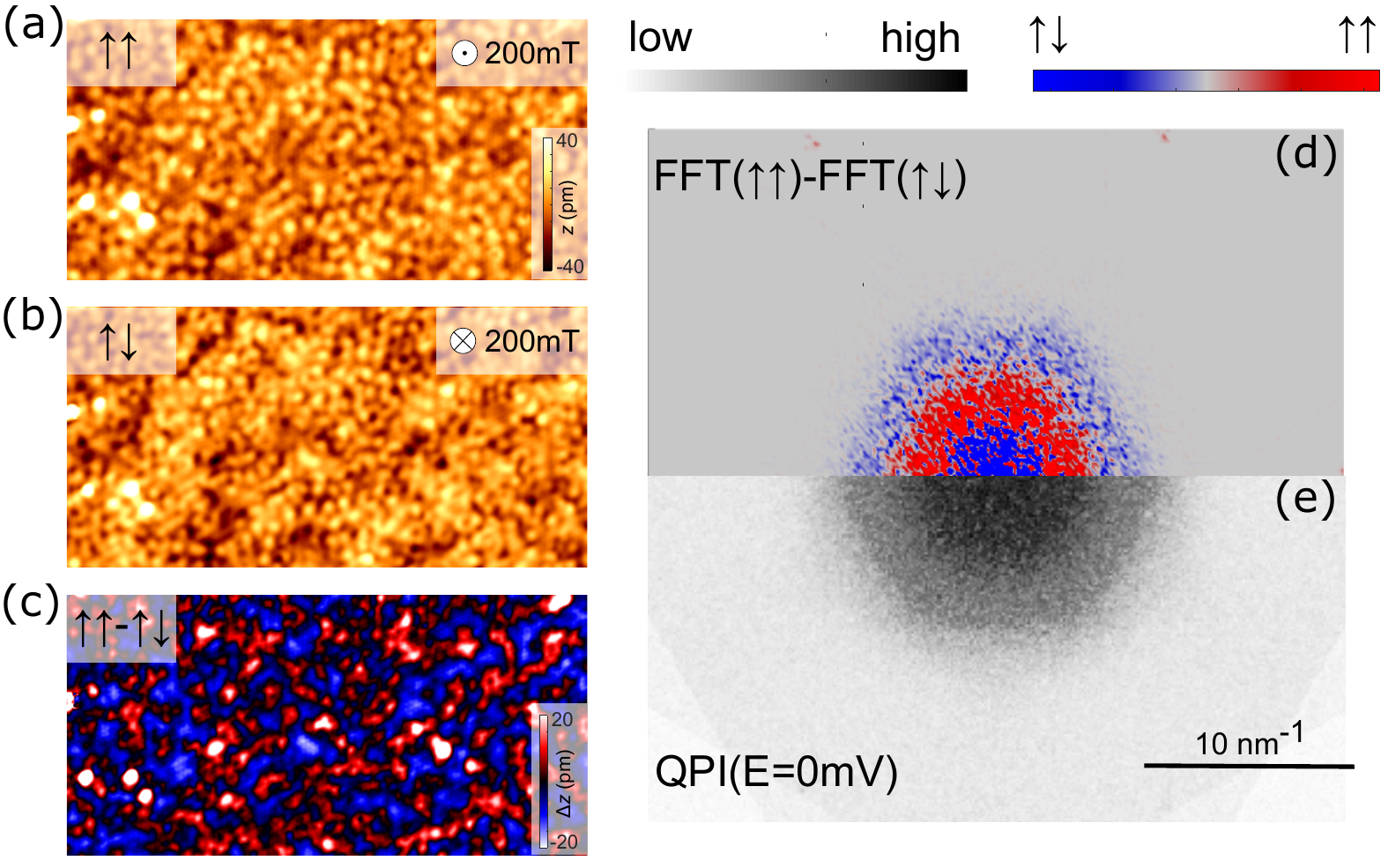}
    \caption{\textbf{Spin-polarized Quasi-particle interference.} (a) SP-STM image ($30\mathrm{mV}$, $100\mathrm{pA}$, $41\times20 \mathrm{nm}^2$) recorded with the tip polarized parallel to the sample by applying a $200\mathrm{mT}$ field along the sample $c$ axis. (b) Image of the same location with the same tip after the magnetization of the tip has been reversed relative to the sample by applying a $200\mathrm{mT}$ field anti-parallel to the sample $c$ axis. Slight differences in the topographic contrast can be observed. (c) The difference of (a) and (b). Sub-surface \ce{Fe} clusters become apparent in the difference image. (d) The quasi-particle interference pattern recorded at the Fermi level ($V=0\mathrm{V}$). (e) Difference of the Fourier transform of topographies ($V_\mathrm{s}=30\mathrm{mV}$, $I_\mathrm{s}=100\mathrm{pA}$) recorded with opposite relative tip and sample spin orientations. Red vectors show up more intensely when the magnetization of tip and sample are parallel, and blue regions when they are antiparallel with respect to each other, showing the different spin-character of the bands. }
    \label{fig:SPfig}
\end{figure}

Using a spin-polarized tip allows to determine the spin character of the different bands. To this end, we have undertaken spin-polarized topographic imaging as a function of applied field. The relative orientation of the tip and sample magnetizations will switch at different fields (see Fig. \ref{fig:SPfig}(a-c)), enabling imaging of the quasi-particle scattering with parallel and anti-parallel alignment of their magnetizations. We observe a significant change in the topographic contrast when the tip and sample are not magnetized in the same direction anymore. Taking the difference of the Fourier transforms of topographies with parallel and antiparallel magnetization (Fig.~\ref{fig:SPfig}(d)) reveals strong changes in the relative intensity of the scattering vectors. By comparing the Fourier transform with the quasi particle interference mapping at the Fermi level (Fig. \ref{fig:SPfig}(e)) we determine that the observed magnetic contrast is due to the spin-dependent imaging of Friedel oscillations between differently polarized bands at the Fermi level and thus reveals information about the spin polarization of the bands in the vicinity of the Fermi energy. 

\section{Discussion}
Our study of \ce{Fe3GeTe2} provides new insights from relating microscopic information obtained from low temperature scanning tunneling microscopy to bulk properties obtained from neutron scattering. We find that in the surface layer, the magnetic properties deviate slightly from those in the bulk, as might be expected from the 3D nature of some parts of the electronic structure. Comparison of the imaging of domain walls with the magnetic exchange interaction $J$ and anisotropy $K$ obtained from neutron scattering reveal a surprisingly good agreement. One would expect the ratio $K/J$ to be slightly larger for the surface layer compared to the bulk due to a larger magnetic anisotropy and smaller exchange coupling: the smaller coordination number of atoms in the surface layer will lead to a decreased exchange energy $J$, while at the same time the lower symmetry is expected to result in a larger anisotropy $K$. It is well known from a number of systems that the magnetic anisotropy increases with reduced dimensionality \cite{gambardella_giant_2003}. The reduction in $J$ in the surface layer is consistent with the lower Curie temperature $T_\mathrm{C}$ found in thin films and in the monolayer limit, where $T_\mathrm C$ is suppressed significantly \cite{Deng2018,Fei_2021}. This system therefore confirms the previously observed trend that while the surface of a magnetic bulk material does not exhibit radically different magnetic properties, there are subtle differences \cite{trainer_magnetic_2021}. 

From band structure calculations, we find that the Fermi surface is dominated by bands of spin majority character with significant $k_z$-dispersion and quasi-2D bands of spin minority character. A comparison with quasi-particle interference imaging reveals dominant wave vectors which are broadly consistent with the band structure calculations. We use spin-polarized imaging to determine the spin-polarization of the bands near the Fermi energy, and find, consistent with the calculations, a hexagonal ring of scattering which we attribute to the bands with spin-majority character embedded in a rather broad distribution of electronic states with opposite spin character.

\section{Acknowledgments}
CT and PW acknowledge funding through EP/R031924/1 and EP/T031441/1, LCR through the Royal Commission for the Exhibition of 1851, IB through the International Max Planck Research School for Chemistry and Physics of Quantum Materials, and HL through the ISIS facility development studentship programme. Access to MACS was provided by the Center for High Resolution Neutron Scattering, a partnership between the National Institute of Standards and Technology and the National Science Foundation under Agreement No. DMR-1508249.

\textbf{Author contributions:} CT, ORA and IB carried out STM measurements, CT and ORA analyzed the STM data. HL and CT derived the expressions for the domain wall width and profile. LCR performed DFT calculations. HL, EC, JARR and CS undertook neutron measurements and crystal growth. PW and CS initiated and led the project. All authors contributed to and discussed the manuscript.

\textbf{Corresponding author}\\
Correspondence should be sent to P. Wahl (\href{mailto:wahl@st-andrews.ac.uk}{wahl@st-andrews.ac.uk}).

\textbf{Competing interests:} The authors declare no competing financial interests.

\textbf{Data Availability:} Underpinning data will be made available at http://$<$DOI to repository will be provided upon acceptance$>$.

\section{Methods}
\subsection{Scanning Tunneling Microscopy}
Scanning Tunneling Microscopy (STM) measurements were undertaken using two home-built STMs \cite{white_stiff_2011,trainer_cryogenic_2017} operating at temperatures down to $1.5\mathrm{K}$ in cryogenic vacuum. Samples are cleaved at a temperature of $\sim 20\mathrm K$ before inserting them in to the STM head.
The spin-polarized STM tips used in this study were prepared using two methods: the first method involved collecting magnetic material from the surface of a sample of an \ce{Fe_{1+x}Te} sample \cite{tipprep,Enayat653} before inserting the \ce{Fe3GeTe2} into the STM. The second method involved picking up ferromagnetic material directly from the \ce{Fe3GeTe2} surface. 
\subsection{Crystal growth}
Single crystals of \ce{Fe3GeTe2} were synthesized using the chemical vapor transport technique.  Sealed quartz ampoules with an outer diameter of 18 mm and inner diameter of 16 mm were loaded with iron, germanium, and tellurium in stoichiometric quantities.  Approximately 10mg of Iodine was loaded as a transport agent.  The iron, germanium, and tellurium powder was initially pumped to $\sim$ 10$^{-5}\mathrm{Torr}$ using a turbo pump to ensure dryness before the iodine was loaded.  The combined reagents were then chilled and pumped to $5\cdot 10^{-3}\mathrm{Torr}$ using an oil-based mechanical pump to avoid damage to the blades of the turbo pump.  The tubes were sealed to be a length of $\sim 15\mathrm{cm}$ and put into a 3-zone furnace such that one end was at $750 ^{\circ}\mathrm{C}$ and the other at $700 ^{\circ}\mathrm{C}$.  A chiller was used to further cool one end of the furnace to increase the temperature gradient.  The temperature gradient was initially inverted for 12 hours to clean one end of the ampoules.  The ampoules were then removed at high temperature with one end cooled using compressed air on removal from the three-zone furnace. The growth resulted in a variety of crystal sizes up to a maximum of the lateral size of $5\times 5 \mathrm{mm}^2$, with a thickness of up to $1\mathrm{mm}$.
\subsection{Neutron scattering}
Neutron scattering measurements were performed on the MACS cold triple axis spectrometer (NIST, Gaithersburg) \cite{MACS}. Single crystals of \ce{Fe3GeTe2} with varying size were edge-aligned on aluminum plates, using the hexagonal morphology as a guide. A coating of hydrogen-free Fomblin oil was applied to the crystals to attach them to the mount and prevent degradation in air. The sample mount was positioned such that the (HHL) reflections lay in the horizontal scattering plane. The sample was then cooled to $T = 5\mathrm{K}$ in a $100\mathrm{mm}$ Orange Cryostat. $E_f$ was fixed to $3.7\mathrm{meV}$ by 20 PG(002) double bounce analyzers and $E_{i}$ was varied between $4.3\mathrm{meV}$ and $\mathrm{13.7}\mathrm{meV}$ by a vertically focused PG(002) monochromator, giving access to energy transfers between $0.5\mathrm{meV}$ and $10\mathrm{meV}$. Cooled \ce{BeO} filters were placed on the scattered side of the sample to remove contamination from higher order scattering. This experimental configuration allowed for an energy resolution of approximately $0.17\mathrm{meV}$ (FWHM) at the elastic line. A two-dimensional map of the scattering intensity along $(H,H,-2)$ was constructed by integrating over $c^{*}$ in a $0.3\mathrm{r.l.u.}$ window about $L = -2$ (Fig.~\ref{magneticcontrast}(a)).

The single-crystal neutron diffraction experiment was performed on hot neutron four-circle diffractometer D9, at ILL \cite{D9doi}. Cu(220) incident-beam monochromator selected a wavelength $\lambda=0.836$ \AA , allowing measurements of Bragg peaks at high momentum transfer. A single-crystal of \ce{Fe3GeTe2} of approximate dimensions $3\times 3\times 2\mathrm{mm}^{3}$ was characterized at four temperatures: $300\mathrm{K}$ (paramagnetic phase), $98\mathrm{K}$, $60\mathrm{K}$ and $30\mathrm{K}$ (ferromagnetic phase). The refinement of nuclear and magnetic structure was performed using \textsc{Fullprof} \cite{fullprof} giving an iron stoichiometry $3-y=2.86(3)$ (from $60\mathrm{K}$ data), and a ferromagnetic moment $\mu=1.6(2)\mu_\mathrm{B}$ at $30\mathrm{K}$. Further details on the refinement are provided in the Supplementary Material.

\subsection{DFT calculations}
DFT calculations to simulate STM images of defects were performed using the Density Functional Theory package Quantum Espresso using Projected Augmented Wavefunctions (PAW) and the PBE exchange-correlation functional. We took a $3\times 3$ supercell of a monolayer of \ce{Fe3GeTe2} with $15\mathrm{\AA}$ vacuum and removed an atom to generate the defect. We then performed structural relaxations on the ferromagnetic 3x3 supercells until the total force was less than $1.0\cdot10^{-3}\mathrm{Ry}/\mathrm{Bohr}$ and the total energy change was less than $1.0\cdot10^{-4}\mathrm{Ry}$. For the structural optimisation we used a $k$-grid of $4\times 4\times 2$. The kinetic energy cutoff was $50\mathrm{Ry}$ for the wavefunctions and $400\mathrm{Ry}$ for the charge density. Using these relaxed structural parameters we then performed a self consistent calculation using a denser k-grid of $8\times 8\times 1$. The total magnetic moment was found to be $2.123 \mu_\mathrm{B}$/Fe for the supercell without a defect, $2.132 \mu_\mathrm{B}$/Fe with a Te defect and $2.004 \mu_\mathrm{B}$/Fe with an Fe defect. Finally, the local density of states was obtained using a $k$-grid of $12\times 12\times 1$ for a bias setpoint of $+100\mathrm{meV}$ ($+0.00735\mathrm{Ry}$) and the STM image at constant current was generated using the CRITIC2 software \cite{Critic2_Software}. For calculations of the bulk electronic structure we used a kinetic energy cutoff of $60\mathrm{Ry}$ for the wavefunctions and $300\mathrm{Ry}$ for the charge density with a $k$-grid of $8\times 8\times 8$.


\begin{thebibliography}{33}
\expandafter\ifx\csname natexlab\endcsname\relax\def\natexlab#1{#1}\fi
\expandafter\ifx\csname bibnamefont\endcsname\relax
  \def\bibnamefont#1{#1}\fi
\expandafter\ifx\csname bibfnamefont\endcsname\relax
  \def\bibfnamefont#1{#1}\fi
\expandafter\ifx\csname citenamefont\endcsname\relax
  \def\citenamefont#1{#1}\fi
\expandafter\ifx\csname url\endcsname\relax
  \def\url#1{\texttt{#1}}\fi
\expandafter\ifx\csname urlprefix\endcsname\relax\def\urlprefix{URL }\fi
\providecommand{\bibinfo}[2]{#2}
\providecommand{\eprint}[2][]{\url{#2}}

\bibitem[{\citenamefont{He et~al.}(2021)\citenamefont{He, Li, Bandyopadhyay,
  and Frauenheim}}]{He_2021}
\bibinfo{author}{\bibfnamefont{J.}~\bibnamefont{He}},
  \bibinfo{author}{\bibfnamefont{S.}~\bibnamefont{Li}},
  \bibinfo{author}{\bibfnamefont{A.}~\bibnamefont{Bandyopadhyay}},
  \bibnamefont{and}
  \bibinfo{author}{\bibfnamefont{T.}~\bibnamefont{Frauenheim}},
  \bibinfo{journal}{ACS NANO LETTERS} \textbf{\bibinfo{volume}{21}},
  \bibinfo{pages}{3237} (\bibinfo{year}{2021}),
  \urlprefix\url{https://doi.org/10.1021/acs.nanolett.1c00520}.

\bibitem[{\citenamefont{Gong et~al.}(2017)\citenamefont{Gong, Li, Li, Ji,
  Stern, Xia, Cao, Bao, Wang, Wang et~al.}}]{Gong2017}
\bibinfo{author}{\bibfnamefont{C.}~\bibnamefont{Gong}},
  \bibinfo{author}{\bibfnamefont{L.}~\bibnamefont{Li}},
  \bibinfo{author}{\bibfnamefont{Z.}~\bibnamefont{Li}},
  \bibinfo{author}{\bibfnamefont{H.}~\bibnamefont{Ji}},
  \bibinfo{author}{\bibfnamefont{A.}~\bibnamefont{Stern}},
  \bibinfo{author}{\bibfnamefont{Y.}~\bibnamefont{Xia}},
  \bibinfo{author}{\bibfnamefont{T.}~\bibnamefont{Cao}},
  \bibinfo{author}{\bibfnamefont{W.}~\bibnamefont{Bao}},
  \bibinfo{author}{\bibfnamefont{C.}~\bibnamefont{Wang}},
  \bibinfo{author}{\bibfnamefont{Y.}~\bibnamefont{Wang}}, \bibnamefont{et~al.},
  \bibinfo{journal}{Nature} \textbf{\bibinfo{volume}{546}},
  \bibinfo{pages}{265} (\bibinfo{year}{2017}), ISSN \bibinfo{issn}{1476-4687},
  \urlprefix\url{https://doi.org/10.1038/nature22060}.

\bibitem[{\citenamefont{Huang et~al.}(2017)\citenamefont{Huang, Clark,
  Navarro-Moratalla, Klein, Cheng, Seyler, Zhong, Schmidgall, McGuire, Cobden
  et~al.}}]{Huang2017}
\bibinfo{author}{\bibfnamefont{B.}~\bibnamefont{Huang}},
  \bibinfo{author}{\bibfnamefont{G.}~\bibnamefont{Clark}},
  \bibinfo{author}{\bibfnamefont{E.}~\bibnamefont{Navarro-Moratalla}},
  \bibinfo{author}{\bibfnamefont{D.~R.} \bibnamefont{Klein}},
  \bibinfo{author}{\bibfnamefont{R.}~\bibnamefont{Cheng}},
  \bibinfo{author}{\bibfnamefont{K.~L.} \bibnamefont{Seyler}},
  \bibinfo{author}{\bibfnamefont{D.}~\bibnamefont{Zhong}},
  \bibinfo{author}{\bibfnamefont{E.}~\bibnamefont{Schmidgall}},
  \bibinfo{author}{\bibfnamefont{M.~A.} \bibnamefont{McGuire}},
  \bibinfo{author}{\bibfnamefont{D.~H.} \bibnamefont{Cobden}},
  \bibnamefont{et~al.}, \bibinfo{journal}{Nature}
  \textbf{\bibinfo{volume}{546}}, \bibinfo{pages}{270} (\bibinfo{year}{2017}),
  ISSN \bibinfo{issn}{1476-4687},
  \urlprefix\url{https://doi.org/10.1038/nature22391}.

\bibitem[{\citenamefont{Fei et~al.}(2018)\citenamefont{Fei, Huang, Malinowski,
  Wang, Song, Sanchez, Yao, Xiao, Zhu, May et~al.}}]{Fei_2021}
\bibinfo{author}{\bibfnamefont{Z.}~\bibnamefont{Fei}},
  \bibinfo{author}{\bibfnamefont{B.}~\bibnamefont{Huang}},
  \bibinfo{author}{\bibfnamefont{P.}~\bibnamefont{Malinowski}},
  \bibinfo{author}{\bibfnamefont{W.}~\bibnamefont{Wang}},
  \bibinfo{author}{\bibfnamefont{T.}~\bibnamefont{Song}},
  \bibinfo{author}{\bibfnamefont{J.}~\bibnamefont{Sanchez}},
  \bibinfo{author}{\bibfnamefont{W.}~\bibnamefont{Yao}},
  \bibinfo{author}{\bibfnamefont{D.}~\bibnamefont{Xiao}},
  \bibinfo{author}{\bibfnamefont{X.}~\bibnamefont{Zhu}},
  \bibinfo{author}{\bibfnamefont{A.~F.} \bibnamefont{May}},
  \bibnamefont{et~al.}, \bibinfo{journal}{Nat. Mater.}
  \textbf{\bibinfo{volume}{17}}, \bibinfo{pages}{778} (\bibinfo{year}{2018}),
  \urlprefix\url{https://doi.org/10.1038/s41563-018-0149-7}.

\bibitem[{\citenamefont{Deng et~al.}(2018)\citenamefont{Deng, Yu, Song, Zhang,
  Wang, Sun, Yi, Wu, Wu, Zhu et~al.}}]{Deng2018}
\bibinfo{author}{\bibfnamefont{Y.}~\bibnamefont{Deng}},
  \bibinfo{author}{\bibfnamefont{Y.}~\bibnamefont{Yu}},
  \bibinfo{author}{\bibfnamefont{Y.}~\bibnamefont{Song}},
  \bibinfo{author}{\bibfnamefont{J.}~\bibnamefont{Zhang}},
  \bibinfo{author}{\bibfnamefont{N.~Z.} \bibnamefont{Wang}},
  \bibinfo{author}{\bibfnamefont{Z.}~\bibnamefont{Sun}},
  \bibinfo{author}{\bibfnamefont{Y.}~\bibnamefont{Yi}},
  \bibinfo{author}{\bibfnamefont{Y.~Z.} \bibnamefont{Wu}},
  \bibinfo{author}{\bibfnamefont{S.}~\bibnamefont{Wu}},
  \bibinfo{author}{\bibfnamefont{J.}~\bibnamefont{Zhu}}, \bibnamefont{et~al.},
  \bibinfo{journal}{Nature} \textbf{\bibinfo{volume}{563}}, \bibinfo{pages}{94}
  (\bibinfo{year}{2018}), ISSN \bibinfo{issn}{1476-4687},
  \urlprefix\url{https://doi.org/10.1038/s41586-018-0626-9}.

\bibitem[{\citenamefont{Deiseroth et~al.}(2006)\citenamefont{Deiseroth,
  Aleksandrov, Reiner, Kienle, and Kremer}}]{deiseroth_fe3gete2_2006}
\bibinfo{author}{\bibfnamefont{H.-J.} \bibnamefont{Deiseroth}},
  \bibinfo{author}{\bibfnamefont{K.}~\bibnamefont{Aleksandrov}},
  \bibinfo{author}{\bibfnamefont{C.}~\bibnamefont{Reiner}},
  \bibinfo{author}{\bibfnamefont{L.}~\bibnamefont{Kienle}}, \bibnamefont{and}
  \bibinfo{author}{\bibfnamefont{R.~K.} \bibnamefont{Kremer}},
  \bibinfo{journal}{European Journal of Inorganic Chemistry}
  \textbf{\bibinfo{volume}{2006}} (\bibinfo{year}{2006}), ISSN
  \bibinfo{issn}{1099-0682},
  \urlprefix\url{https://onlinelibrary.wiley.com/doi/abs/10.1002/ejic.200501020}.

\bibitem[{\citenamefont{May et~al.}(2016)\citenamefont{May, Calder, Cantoni,
  Cao, and McGuire}}]{may_magnetic_2016}
\bibinfo{author}{\bibfnamefont{A.~F.} \bibnamefont{May}},
  \bibinfo{author}{\bibfnamefont{S.}~\bibnamefont{Calder}},
  \bibinfo{author}{\bibfnamefont{C.}~\bibnamefont{Cantoni}},
  \bibinfo{author}{\bibfnamefont{H.}~\bibnamefont{Cao}}, \bibnamefont{and}
  \bibinfo{author}{\bibfnamefont{M.~A.} \bibnamefont{McGuire}},
  \bibinfo{journal}{Phys. Rev. B} \textbf{\bibinfo{volume}{93}},
  \bibinfo{pages}{014411} (\bibinfo{year}{2016}), ISSN
  \bibinfo{issn}{2469-9950, 2469-9969},
  \urlprefix\url{https://link.aps.org/doi/10.1103/PhysRevB.93.014411}.

\bibitem[{\citenamefont{Zhang et~al.}(2018)\citenamefont{Zhang, Lu, Zhu, Tan,
  Feng, Liu, Zhang, Chen, Liu, Luo et~al.}}]{zhang_emergence_2018}
\bibinfo{author}{\bibfnamefont{Y.}~\bibnamefont{Zhang}},
  \bibinfo{author}{\bibfnamefont{H.}~\bibnamefont{Lu}},
  \bibinfo{author}{\bibfnamefont{X.}~\bibnamefont{Zhu}},
  \bibinfo{author}{\bibfnamefont{S.}~\bibnamefont{Tan}},
  \bibinfo{author}{\bibfnamefont{W.}~\bibnamefont{Feng}},
  \bibinfo{author}{\bibfnamefont{Q.}~\bibnamefont{Liu}},
  \bibinfo{author}{\bibfnamefont{W.}~\bibnamefont{Zhang}},
  \bibinfo{author}{\bibfnamefont{Q.}~\bibnamefont{Chen}},
  \bibinfo{author}{\bibfnamefont{Y.}~\bibnamefont{Liu}},
  \bibinfo{author}{\bibfnamefont{X.}~\bibnamefont{Luo}}, \bibnamefont{et~al.},
  \bibinfo{journal}{Sci. Adv.} \textbf{\bibinfo{volume}{4}},
  \bibinfo{pages}{eaao6791} (\bibinfo{year}{2018}), ISSN
  \bibinfo{issn}{2375-2548},
  \urlprefix\url{http://advances.sciencemag.org/lookup/doi/10.1126/sciadv.aao6791}.

\bibitem[{\citenamefont{Tian et~al.}(2019)\citenamefont{Tian, Wang, Ji, Wang,
  Xia, Wang, Liu, Zhang, and Cheng}}]{Tian2019}
\bibinfo{author}{\bibfnamefont{C.-K.} \bibnamefont{Tian}},
  \bibinfo{author}{\bibfnamefont{C.}~\bibnamefont{Wang}},
  \bibinfo{author}{\bibfnamefont{W.}~\bibnamefont{Ji}},
  \bibinfo{author}{\bibfnamefont{J.-C.} \bibnamefont{Wang}},
  \bibinfo{author}{\bibfnamefont{T.-L.} \bibnamefont{Xia}},
  \bibinfo{author}{\bibfnamefont{L.}~\bibnamefont{Wang}},
  \bibinfo{author}{\bibfnamefont{J.-J.} \bibnamefont{Liu}},
  \bibinfo{author}{\bibfnamefont{H.-X.} \bibnamefont{Zhang}}, \bibnamefont{and}
  \bibinfo{author}{\bibfnamefont{P.}~\bibnamefont{Cheng}},
  \bibinfo{journal}{Phys. Rev. B} \textbf{\bibinfo{volume}{99}},
  \bibinfo{pages}{184428} (\bibinfo{year}{2019}),
  \urlprefix\url{https://link.aps.org/doi/10.1103/PhysRevB.99.184428}.

\bibitem[{\citenamefont{Kim et~al.}(2018)\citenamefont{Kim, Seo, Lee, Ko, Kim,
  Jang, Ok, Lee, Jo, Kang et~al.}}]{Kim2018}
\bibinfo{author}{\bibfnamefont{K.}~\bibnamefont{Kim}},
  \bibinfo{author}{\bibfnamefont{J.}~\bibnamefont{Seo}},
  \bibinfo{author}{\bibfnamefont{E.}~\bibnamefont{Lee}},
  \bibinfo{author}{\bibfnamefont{K.-T.} \bibnamefont{Ko}},
  \bibinfo{author}{\bibfnamefont{B.~S.} \bibnamefont{Kim}},
  \bibinfo{author}{\bibfnamefont{B.~G.} \bibnamefont{Jang}},
  \bibinfo{author}{\bibfnamefont{J.~M.} \bibnamefont{Ok}},
  \bibinfo{author}{\bibfnamefont{J.}~\bibnamefont{Lee}},
  \bibinfo{author}{\bibfnamefont{Y.~J.} \bibnamefont{Jo}},
  \bibinfo{author}{\bibfnamefont{W.}~\bibnamefont{Kang}}, \bibnamefont{et~al.},
  \bibinfo{journal}{Nature Materials} \textbf{\bibinfo{volume}{17}},
  \bibinfo{pages}{794} (\bibinfo{year}{2018}), ISSN \bibinfo{issn}{1476-4660},
  \urlprefix\url{https://doi.org/10.1038/s41563-018-0132-3}.

\bibitem[{\citenamefont{Lovesey}(1984)}]{Lovesey}
\bibinfo{author}{\bibfnamefont{S.~W.} \bibnamefont{Lovesey}},
  \emph{\bibinfo{title}{Theory of neutron scattering from condensed matter}}
  (\bibinfo{publisher}{Clarendon Press}, \bibinfo{address}{United Kingdom},
  \bibinfo{year}{1984}), ISBN \bibinfo{isbn}{0-19-852017-4}.

\bibitem[{\citenamefont{Owerre}(2016)}]{Owerre}
\bibinfo{author}{\bibfnamefont{S.~A.} \bibnamefont{Owerre}},
  \bibinfo{journal}{Journal of Physics: Condensed Matter}
  \textbf{\bibinfo{volume}{28}}, \bibinfo{pages}{386001}
  (\bibinfo{year}{2016}), ISSN \bibinfo{issn}{0953-8984, 1361-648X}.

\bibitem[{\citenamefont{Pershoguba et~al.}(2018)\citenamefont{Pershoguba,
  Banerjee, Lashley, Park, \AA{}gren, Aeppli, and Balatsky}}]{Pershoguba}
\bibinfo{author}{\bibfnamefont{S.~S.} \bibnamefont{Pershoguba}},
  \bibinfo{author}{\bibfnamefont{S.}~\bibnamefont{Banerjee}},
  \bibinfo{author}{\bibfnamefont{J.~C.} \bibnamefont{Lashley}},
  \bibinfo{author}{\bibfnamefont{J.}~\bibnamefont{Park}},
  \bibinfo{author}{\bibfnamefont{H.}~\bibnamefont{\AA{}gren}},
  \bibinfo{author}{\bibfnamefont{G.}~\bibnamefont{Aeppli}}, \bibnamefont{and}
  \bibinfo{author}{\bibfnamefont{A.~V.} \bibnamefont{Balatsky}},
  \bibinfo{journal}{Phys. Rev. X} \textbf{\bibinfo{volume}{8}},
  \bibinfo{pages}{011010} (\bibinfo{year}{2018}),
  \urlprefix\url{https://link.aps.org/doi/10.1103/PhysRevX.8.011010}.

\bibitem[{\citenamefont{Tan et~al.}(2018)\citenamefont{Tan, Lee, Jung, Park,
  Albarakati, Partridge, Field, McCulloch, Wang, and Lee}}]{Tan2018}
\bibinfo{author}{\bibfnamefont{C.}~\bibnamefont{Tan}},
  \bibinfo{author}{\bibfnamefont{J.}~\bibnamefont{Lee}},
  \bibinfo{author}{\bibfnamefont{S.-G.} \bibnamefont{Jung}},
  \bibinfo{author}{\bibfnamefont{T.}~\bibnamefont{Park}},
  \bibinfo{author}{\bibfnamefont{S.}~\bibnamefont{Albarakati}},
  \bibinfo{author}{\bibfnamefont{J.}~\bibnamefont{Partridge}},
  \bibinfo{author}{\bibfnamefont{M.~R.} \bibnamefont{Field}},
  \bibinfo{author}{\bibfnamefont{D.~G.} \bibnamefont{McCulloch}},
  \bibinfo{author}{\bibfnamefont{L.}~\bibnamefont{Wang}}, \bibnamefont{and}
  \bibinfo{author}{\bibfnamefont{C.}~\bibnamefont{Lee}},
  \bibinfo{journal}{Nature Communications} \textbf{\bibinfo{volume}{9}},
  \bibinfo{pages}{1554} (\bibinfo{year}{2018}), ISSN \bibinfo{issn}{2041-1723},
  \urlprefix\url{https://doi.org/10.1038/s41467-018-04018-w}.

\bibitem[{\citenamefont{Calder et~al.}(2019)\citenamefont{Calder, Kolesnikov,
  and May}}]{exchangecouple}
\bibinfo{author}{\bibfnamefont{S.}~\bibnamefont{Calder}},
  \bibinfo{author}{\bibfnamefont{A.~I.} \bibnamefont{Kolesnikov}},
  \bibnamefont{and} \bibinfo{author}{\bibfnamefont{A.~F.} \bibnamefont{May}},
  \bibinfo{journal}{Phys. Rev. B} \textbf{\bibinfo{volume}{99}},
  \bibinfo{pages}{094423} (\bibinfo{year}{2019}),
  \urlprefix\url{https://link.aps.org/doi/10.1103/PhysRevB.99.094423}.

\bibitem[{\citenamefont{Wiesendanger}(2009)}]{Wiesendangerrev}
\bibinfo{author}{\bibfnamefont{R.}~\bibnamefont{Wiesendanger}},
  \bibinfo{journal}{Rev. Mod. Phys.} \textbf{\bibinfo{volume}{81}},
  \bibinfo{pages}{1495} (\bibinfo{year}{2009}),
  \urlprefix\url{https://link.aps.org/doi/10.1103/RevModPhys.81.1495}.

\bibitem[{\citenamefont{Bode}(2003)}]{Bode_2003}
\bibinfo{author}{\bibfnamefont{M.}~\bibnamefont{Bode}},
  \bibinfo{journal}{Reports on Progress in Physics}
  \textbf{\bibinfo{volume}{66}}, \bibinfo{pages}{523} (\bibinfo{year}{2003}),
  \urlprefix\url{https://doi.org/10.1088/0034-4885/66/4/203}.

\bibitem[{\citenamefont{Moreno et~al.}(2016)\citenamefont{Moreno, Evans,
  Khmelevskyi, Mu\~noz, Chantrell, and Chubykalo-Fesenko}}]{hexagonaldelta}
\bibinfo{author}{\bibfnamefont{R.}~\bibnamefont{Moreno}},
  \bibinfo{author}{\bibfnamefont{R.~F.~L.} \bibnamefont{Evans}},
  \bibinfo{author}{\bibfnamefont{S.}~\bibnamefont{Khmelevskyi}},
  \bibinfo{author}{\bibfnamefont{M.~C.} \bibnamefont{Mu\~noz}},
  \bibinfo{author}{\bibfnamefont{R.~W.} \bibnamefont{Chantrell}},
  \bibnamefont{and}
  \bibinfo{author}{\bibfnamefont{O.}~\bibnamefont{Chubykalo-Fesenko}},
  \bibinfo{journal}{Phys. Rev. B} \textbf{\bibinfo{volume}{94}},
  \bibinfo{pages}{104433} (\bibinfo{year}{2016}),
  \urlprefix\url{https://link.aps.org/doi/10.1103/PhysRevB.94.104433}.

\bibitem[{\citenamefont{Aharoni}(2000)}]{aharoni}
\bibinfo{author}{\bibfnamefont{A.}~\bibnamefont{Aharoni}},
  \emph{\bibinfo{title}{Introduction to the Theory of Ferromagnetism}},
  International Series of Monographs on Physics (\bibinfo{publisher}{Clarendon
  Press}, \bibinfo{year}{2000}), ISBN \bibinfo{isbn}{9780198508090},
  \urlprefix\url{https://books.google.co.uk/books?id=Ru-z9b3WcfMC}.

\bibitem[{\citenamefont{Berbil-Bautista
  et~al.}(2007)\citenamefont{Berbil-Bautista, Krause, Bode, and
  Wiesendanger}}]{domain1}
\bibinfo{author}{\bibfnamefont{L.}~\bibnamefont{Berbil-Bautista}},
  \bibinfo{author}{\bibfnamefont{S.}~\bibnamefont{Krause}},
  \bibinfo{author}{\bibfnamefont{M.}~\bibnamefont{Bode}}, \bibnamefont{and}
  \bibinfo{author}{\bibfnamefont{R.}~\bibnamefont{Wiesendanger}},
  \bibinfo{journal}{Phys. Rev. B} \textbf{\bibinfo{volume}{76}},
  \bibinfo{pages}{064411} (\bibinfo{year}{2007}),
  \urlprefix\url{https://link.aps.org/doi/10.1103/PhysRevB.76.064411}.

\bibitem[{\citenamefont{Ravli\ifmmode~\acute{c}\else \'{c}\fi{}
  et~al.}(2003)\citenamefont{Ravli\ifmmode~\acute{c}\else \'{c}\fi{}, Bode,
  Kubetzka, and Wiesendanger}}]{domain2}
\bibinfo{author}{\bibfnamefont{R.}~\bibnamefont{Ravli\ifmmode~\acute{c}\else
  \'{c}\fi{}}}, \bibinfo{author}{\bibfnamefont{M.}~\bibnamefont{Bode}},
  \bibinfo{author}{\bibfnamefont{A.}~\bibnamefont{Kubetzka}}, \bibnamefont{and}
  \bibinfo{author}{\bibfnamefont{R.}~\bibnamefont{Wiesendanger}},
  \bibinfo{journal}{Phys. Rev. B} \textbf{\bibinfo{volume}{67}},
  \bibinfo{pages}{174411} (\bibinfo{year}{2003}),
  \urlprefix\url{https://link.aps.org/doi/10.1103/PhysRevB.67.174411}.

\bibitem[{\citenamefont{Blundell}(2001)}]{Blundell}
\bibinfo{author}{\bibfnamefont{S.}~\bibnamefont{Blundell}},
  \emph{\bibinfo{title}{Magnetism in condensed matter}}
  (\bibinfo{publisher}{Oxford ; New York : Oxford University Press, 2001.},
  \bibinfo{year}{2001}), \bibinfo{note}{includes bibliographical references and
  index.},
  \urlprefix\url{https://search.library.wisc.edu/catalog/999921680802121}.

\bibitem[{\citenamefont{Xu et~al.}(2020)\citenamefont{Xu, Li, Duan, Zhang,
  Chen, Kang, Liang, Chen, Xia, Xu et~al.}}]{PRBARPES}
\bibinfo{author}{\bibfnamefont{X.}~\bibnamefont{Xu}},
  \bibinfo{author}{\bibfnamefont{Y.~W.} \bibnamefont{Li}},
  \bibinfo{author}{\bibfnamefont{S.~R.} \bibnamefont{Duan}},
  \bibinfo{author}{\bibfnamefont{S.~L.} \bibnamefont{Zhang}},
  \bibinfo{author}{\bibfnamefont{Y.~J.} \bibnamefont{Chen}},
  \bibinfo{author}{\bibfnamefont{L.}~\bibnamefont{Kang}},
  \bibinfo{author}{\bibfnamefont{A.~J.} \bibnamefont{Liang}},
  \bibinfo{author}{\bibfnamefont{C.}~\bibnamefont{Chen}},
  \bibinfo{author}{\bibfnamefont{W.}~\bibnamefont{Xia}},
  \bibinfo{author}{\bibfnamefont{Y.}~\bibnamefont{Xu}}, \bibnamefont{et~al.},
  \bibinfo{journal}{Phys. Rev. B} \textbf{\bibinfo{volume}{101}},
  \bibinfo{pages}{201104} (\bibinfo{year}{2020}),
  \urlprefix\url{https://link.aps.org/doi/10.1103/PhysRevB.101.201104}.

\bibitem[{\citenamefont{Gambardella}(2003)}]{gambardella_giant_2003}
\bibinfo{author}{\bibfnamefont{P.}~\bibnamefont{Gambardella}},
  \bibinfo{journal}{Science} \textbf{\bibinfo{volume}{300}},
  \bibinfo{pages}{1130} (\bibinfo{year}{2003}), ISSN \bibinfo{issn}{00368075,
  10959203},
  \urlprefix\url{https://www.sciencemag.org/lookup/doi/10.1126/science.1082857}.

\bibitem[{\citenamefont{Trainer et~al.}(2021)\citenamefont{Trainer, Songvilay,
  Qureshi, Stunault, Yim, Rodriguez, Heil, Tsurkan, Green, Loidl
  et~al.}}]{trainer_magnetic_2021}
\bibinfo{author}{\bibfnamefont{C.}~\bibnamefont{Trainer}},
  \bibinfo{author}{\bibfnamefont{M.}~\bibnamefont{Songvilay}},
  \bibinfo{author}{\bibfnamefont{N.}~\bibnamefont{Qureshi}},
  \bibinfo{author}{\bibfnamefont{A.}~\bibnamefont{Stunault}},
  \bibinfo{author}{\bibfnamefont{C.~M.} \bibnamefont{Yim}},
  \bibinfo{author}{\bibfnamefont{E.~E.} \bibnamefont{Rodriguez}},
  \bibinfo{author}{\bibfnamefont{C.}~\bibnamefont{Heil}},
  \bibinfo{author}{\bibfnamefont{V.}~\bibnamefont{Tsurkan}},
  \bibinfo{author}{\bibfnamefont{M.~A.} \bibnamefont{Green}},
  \bibinfo{author}{\bibfnamefont{A.}~\bibnamefont{Loidl}},
  \bibnamefont{et~al.}, \bibinfo{journal}{Phys. Rev. B}
  \textbf{\bibinfo{volume}{103}}, \bibinfo{pages}{024406}
  (\bibinfo{year}{2021}), ISSN \bibinfo{issn}{2469-9950, 2469-9969},
  \urlprefix\url{https://link.aps.org/doi/10.1103/PhysRevB.103.024406}.

\bibitem[{\citenamefont{White et~al.}(2011)\citenamefont{White, Singh, and
  Wahl}}]{white_stiff_2011}
\bibinfo{author}{\bibfnamefont{S.~C.} \bibnamefont{White}},
  \bibinfo{author}{\bibfnamefont{U.~R.} \bibnamefont{Singh}}, \bibnamefont{and}
  \bibinfo{author}{\bibfnamefont{P.}~\bibnamefont{Wahl}},
  \bibinfo{journal}{Review of Scientific Instruments}
  \textbf{\bibinfo{volume}{82}}, \bibinfo{pages}{113708}
  (\bibinfo{year}{2011}),
  \urlprefix\url{http://scitation.aip.org/content/aip/journal/rsi/82/11/10.1063/1.3663611}.

\bibitem[{\citenamefont{Trainer et~al.}(2017)\citenamefont{Trainer, Yim,
  McLaren, and Wahl}}]{trainer_cryogenic_2017}
\bibinfo{author}{\bibfnamefont{C.}~\bibnamefont{Trainer}},
  \bibinfo{author}{\bibfnamefont{C.~M.} \bibnamefont{Yim}},
  \bibinfo{author}{\bibfnamefont{M.}~\bibnamefont{McLaren}}, \bibnamefont{and}
  \bibinfo{author}{\bibfnamefont{P.}~\bibnamefont{Wahl}},
  \bibinfo{journal}{Review of Scientific Instruments}
  \textbf{\bibinfo{volume}{88}}, \bibinfo{pages}{093705}
  (\bibinfo{year}{2017}), ISSN \bibinfo{issn}{0034-6748, 1089-7623},
  \urlprefix\url{http://aip.scitation.org/doi/10.1063/1.4995688}.

\bibitem[{\citenamefont{Singh et~al.}(2015)\citenamefont{Singh, Aluru, Liu,
  Lin, and Wahl}}]{tipprep}
\bibinfo{author}{\bibfnamefont{U.~R.} \bibnamefont{Singh}},
  \bibinfo{author}{\bibfnamefont{R.}~\bibnamefont{Aluru}},
  \bibinfo{author}{\bibfnamefont{Y.}~\bibnamefont{Liu}},
  \bibinfo{author}{\bibfnamefont{C.}~\bibnamefont{Lin}}, \bibnamefont{and}
  \bibinfo{author}{\bibfnamefont{P.}~\bibnamefont{Wahl}},
  \bibinfo{journal}{Phys. Rev. B} \textbf{\bibinfo{volume}{91}},
  \bibinfo{pages}{161111} (\bibinfo{year}{2015}),
  \urlprefix\url{https://link.aps.org/doi/10.1103/PhysRevB.91.161111}.

\bibitem[{\citenamefont{Enayat et~al.}(2014)\citenamefont{Enayat, Sun, Singh,
  Aluru, Schmaus, Yaresko, Liu, Lin, Tsurkan, Loidl et~al.}}]{Enayat653}
\bibinfo{author}{\bibfnamefont{M.}~\bibnamefont{Enayat}},
  \bibinfo{author}{\bibfnamefont{Z.}~\bibnamefont{Sun}},
  \bibinfo{author}{\bibfnamefont{U.~R.} \bibnamefont{Singh}},
  \bibinfo{author}{\bibfnamefont{R.}~\bibnamefont{Aluru}},
  \bibinfo{author}{\bibfnamefont{S.}~\bibnamefont{Schmaus}},
  \bibinfo{author}{\bibfnamefont{A.}~\bibnamefont{Yaresko}},
  \bibinfo{author}{\bibfnamefont{Y.}~\bibnamefont{Liu}},
  \bibinfo{author}{\bibfnamefont{C.}~\bibnamefont{Lin}},
  \bibinfo{author}{\bibfnamefont{V.}~\bibnamefont{Tsurkan}},
  \bibinfo{author}{\bibfnamefont{A.}~\bibnamefont{Loidl}},
  \bibnamefont{et~al.}, \bibinfo{journal}{Science}
  \textbf{\bibinfo{volume}{345}}, \bibinfo{pages}{653} (\bibinfo{year}{2014}),
  ISSN \bibinfo{issn}{0036-8075},
  \eprint{https://science.sciencemag.org/content/345/6197/653.full.pdf},
  \urlprefix\url{https://science.sciencemag.org/content/345/6197/653}.

\bibitem[{\citenamefont{Rodriguez et~al.}(2008)\citenamefont{Rodriguez, Adler,
  Brand, Broholm, Cook, Brocker, Hammond, Huang, Hundertmark, Lynn
  et~al.}}]{MACS}
\bibinfo{author}{\bibfnamefont{J.~A.} \bibnamefont{Rodriguez}},
  \bibinfo{author}{\bibfnamefont{D.~M.} \bibnamefont{Adler}},
  \bibinfo{author}{\bibfnamefont{P.~C.} \bibnamefont{Brand}},
  \bibinfo{author}{\bibfnamefont{C.}~\bibnamefont{Broholm}},
  \bibinfo{author}{\bibfnamefont{J.~C.} \bibnamefont{Cook}},
  \bibinfo{author}{\bibfnamefont{C.}~\bibnamefont{Brocker}},
  \bibinfo{author}{\bibfnamefont{R.}~\bibnamefont{Hammond}},
  \bibinfo{author}{\bibfnamefont{Z.}~\bibnamefont{Huang}},
  \bibinfo{author}{\bibfnamefont{P.}~\bibnamefont{Hundertmark}},
  \bibinfo{author}{\bibfnamefont{J.~W.} \bibnamefont{Lynn}},
  \bibnamefont{et~al.}, \bibinfo{journal}{Measurement Science and Technology}
  \textbf{\bibinfo{volume}{19}}, \bibinfo{pages}{034023}
  (\bibinfo{year}{2008}),
  \urlprefix\url{https://doi.org/10.1088/0957-0233/19/3/034023}.

\bibitem[{\citenamefont{Chris et~al.}()\citenamefont{Chris, Nebil, Edmond,
  Ramon, Harry, Elise, and Navid}}]{D9doi}
\bibinfo{author}{\bibfnamefont{S.}~\bibnamefont{Chris}},
  \bibinfo{author}{\bibfnamefont{A.~K.} \bibnamefont{Nebil}},
  \bibinfo{author}{\bibfnamefont{C.}~\bibnamefont{Edmond}},
  \bibinfo{author}{\bibfnamefont{F.~R.~O.} \bibnamefont{Ramon}},
  \bibinfo{author}{\bibfnamefont{L.}~\bibnamefont{Harry}},
  \bibinfo{author}{\bibfnamefont{P.}~\bibnamefont{Elise}}, \bibnamefont{and}
  \bibinfo{author}{\bibfnamefont{Q.}~\bibnamefont{Navid}},
  \emph{\bibinfo{title}{The nuclear structure of {CVT} grown {Fe3}-{xGeTe2}}},
  \bibinfo{note}{type: dataset},
  \urlprefix\url{https://doi.ill.fr/10.5291/ILL-DATA.5-11-440}.

\bibitem[{\citenamefont{{Rodr{\'i}guez-Carvajal}}(1993)}]{fullprof}
\bibinfo{author}{\bibfnamefont{J.}~\bibnamefont{{Rodr{\'i}guez-Carvajal}}},
  \bibinfo{journal}{Physica B: Condensed Matter}
  \textbf{\bibinfo{volume}{192}}, \bibinfo{pages}{55} (\bibinfo{year}{1993}),
  ISSN \bibinfo{issn}{09214526},
  \urlprefix\url{https://doi.org/10.1016/0921-4526(93)90108-I}.

\bibitem[{\citenamefont{de-la Roza et~al.}(2014)\citenamefont{de-la Roza,
  Johnson, and Luaña}}]{Critic2_Software}
\bibinfo{author}{\bibfnamefont{A.~O.} \bibnamefont{de-la Roza}},
  \bibinfo{author}{\bibfnamefont{E.~R.} \bibnamefont{Johnson}},
  \bibnamefont{and} \bibinfo{author}{\bibfnamefont{V.}~\bibnamefont{Luaña}},
  \bibinfo{journal}{Computer Physics Communications}
  \textbf{\bibinfo{volume}{185}}, \bibinfo{pages}{1007} (\bibinfo{year}{2014}),
  ISSN \bibinfo{issn}{0010-4655},
  \urlprefix\url{https://www.sciencedirect.com/science/article/pii/S0010465513003718}.

\end{thebibliography}



\section*{Supplementary material}

# Supplementary Material for 'Relating spin-polarized STM imaging and inelastic neutron scattering in van-der-Waals ferromagnet $Fe_3GeTe_2$'

Christopher Trainer,[1] Olivia R. Armitage,[1] Harry Lane,[2,3,4] Luke C. Rhodes,[1] Edmond Chan,[2,5] Izidor Benedičič,[1] J. A. Rodriguez-Rivera,[6,7] O. Fabelo,[5] Chris Stock,[2] and Peter Wahl[1]

[1]SUPA, School of Physics and Astronomy,
University of St Andrews, North Haugh,
St Andrews, Fife, KY16 9SS, United Kingdom

[2]School of Physics and Astronomy, University of Edinburgh,
Edinburgh EH9 3JZ, United Kingdom

[3]School of Chemistry and Centre for Science at Extreme Conditions,
University of Edinburgh, Edinburgh EH9 3FJ, United Kingdom

[4]ISIS Pulsed Neutron and Muon Source,
STFC Rutherford Appleton Laboratory, Harwell Campus,
Didcot, Oxon, OX11 0QX, United Kingdom

[5]Institute Laue-Langevin, 6 rue Jules Horowitz,
Boite Postale 156, 38042 Grenoble Cedex 9, France

[6]NIST Center for Neutron Research, National Institute of
Standards and Technology, Gaithersburg, Maryland 20899, USA

[7]Department of Materials Science and Engineering,
University of Maryland, College Park, Maryland 20742, USA

## S1. SINGLE-CRYSTAL NEUTRON DIFFRACTION

A single-crystal of $Fe_{3-y}GeTe_2$ of approximate dimensions $3 \times 3 \times 2$ mm$^3$ was characterized on the hot neutron four-circle diffractometer D9 at ILL [1]. We used a Cu(220) incident-beam monochromator to select a wavelength $\lambda = 0.836$Å, allowing measurements of Bragg peaks at high momentum transfer. Bragg reflections were collected at four temperatures: 300K (paramagnetic phase), 98K, 60K and 30K (ferromagnetic phase).

Both nuclear and magnetic structure were refined using FULLPROF within nuclear space group *P6$_3$/mmc* (No. 194). Scale, extinction parameters, atomic positions and individual isotropic displacement were refined at each temperature, except at 30 K were scale and extinction parameters were fixed, given the lower number of reflections measured. The occupancy of interstitial Fe2 (Wyckoff position 2c) was refined, and an iron stoichiometry $3 - y = 2.86(3)$ was found from the 60K data. The best magnetic refinement was obtained using magnetic space-group *P6$_3$/mm'c'* (No. 194.27), which leads to ferromagnetism along the $c$-axis. The magnetic moments for Fe1 and Fe2 were constrained to be the same, leading to $\mu = 1.6(2)\mu_B$ at 30K. The refined parameters are listed in Tab. S1, and the resulting observed intensities versus calculated ones are shown in Fig. S1.

## S2. FIELD CONTROL OF FERROMAGNETIC DOMAIN WALLS

To investigate the magnetism of the ferromagnetic domains in $Fe_{3-y}GeTe_2$ we control both the surface and tip magnetization using an applied field. Figure S2 shows topographic images of the same position on the sample surface as a function of applied magnetic field. Starting from a negative field, a clear transition is found with increasing field at positive fields when the sample magnetization switches direction. Figure S2(a-d) shows topographic images taken at different fields from (a) 100mT to (d) 250mT in increments of 50mT. Between panels (b) and (d) the magnetization of the sample reverses, with a domain wall crossing the imaged area in panel (c). This can be better seen in difference images, shown in Figure S2(e-g), where the image obtained at 150mT has been subtracted from the ones taken at 100mT, 200mT and 250mT. The difference image Figure S2(e) shows almost no contrast, consistent with the relative orientation of the tip and sample magnetization staying the same. In Figure S2(f) the domain wall crossing at 200mT can be clearly seen, with the sample magnetization in the upper right hand corner of the image having

TABLE S1: Parameters of a $Fe_{3-y}GeTe_2$ single-crystal refined with FULLPROF within nuclear space group $P6_3/mmc$ (No. 194) and magnetic space-group $P6_3/mm'c'$ (No. 194.27).

| $T = 300$ K | | | | | | | Measured reflections, unique, observed ($I > 2\sigma$): 480, 159, 154 |
|---|---|---|---|---|---|---|---|
| $R_{\text{int}} = 1.95\%$ | $R_{\text{Bragg}} = 6.10\%$ | $\chi^2 = 4.53$ | | | | | |
| $a = b = 3.9920(2)$ Å | $c = 16.343(3)$ Å | | | | | | |

| Atoms | Wyckoff | $x$ | $y$ | $z$ | $B_{\text{iso}}$ (Å$^2$) | Occ. | $M_z$ |
|---|---|---|---|---|---|---|---|
| Fe1 | 4e | 0.000 | 0.000 | 0.6715(3) | 0.67(3) | 1 | / |
| Fe2 | 2c | 0.667 | 0.333 | 0.750 | 0.83(7) | 0.86(2) | / |
| Ge | 2d | 0.333 | 0.667 | 0.750 | 1.91(8) | 1 | / |
| Te | 4f | 0.667 | 0.333 | 0.5900(7) | 0.82(4) | 1 | / |

| $T = 98$ K | | | | | | | Measured reflections, unique, observed ($I > 2\sigma$): 382, 128, 126 |
|---|---|---|---|---|---|---|---|
| $R_{\text{int}} = 1.6\%$ | $R_{\text{Bragg}} = 6.22\%$ | $\chi^2 = 7.20$ | | | | | |
| $a = b = 3.9794(5)$ Å | $c = 16.304(3)$ Å | | | | | | |

| Atoms | Wyckoff | $x$ | $y$ | $z$ | $B_{\text{iso}}$ (Å$^2$) | Occ. | $M_z$ |
|---|---|---|---|---|---|---|---|
| Fe1 | 4e | 0.000 | 0.000 | 0.6718(2) | 0.31(3) | 1 | 1.4(3) |
| Fe2 | 2c | 0.667 | 0.333 | 0.750 | 0.63(9) | 0.87(3) | 1.4(3) |
| Ge | 2d | 0.333 | 0.667 | 0.750 | 1.7(1) | 1 | / |
| Te | 4f | 0.667 | 0.333 | 0.5894(3) | 0.35(4) | 1 | / |

| $T = 60$ K | | | | | | | Measured reflections, unique, observed ($I > 2\sigma$): 645, 189, 188 |
|---|---|---|---|---|---|---|---|
| $R_{\text{int}} = 1.52\%$ | $R_{\text{Bragg}} = 8.25\%$ | $\chi^2 = 14.13$ | | | | | |
| $a = b = 3.9785(5)$ Å | $c = 16.286(4)$ Å | | | | | | |

| Atoms | Wyckoff | $x$ | $y$ | $z$ | $B_{\text{iso}}$ (Å$^2$) | Occ. | $M_z$ |
|---|---|---|---|---|---|---|---|
| Fe1 | 4e | 0.000 | 0.000 | 0.6718(3) | 0.24(2) | 1 | 1.5(4) |
| Fe2 | 2c | 0.667 | 0.333 | 0.750 | 0.51(7) | 0.86(3) | 1.5(4) |
| Ge | 2d | 0.333 | 0.667 | 0.750 | 1.57(9) | 1 | / |
| Te | 4f | 0.667 | 0.333 | 0.5896(7) | 0.25(3) | 1 | / |

| $T = 30$ K | | | | | | | Measured reflections, unique, observed ($I > 2\sigma$): 200, 76, 76 |
|---|---|---|---|---|---|---|---|
| $R_{\text{int}} = 1.44\%$ | $R_{\text{Bragg}} = 4.44\%$ | $\chi^2 = 4.04$ | | | | | |
| $a = b = 3.9779(1)$ Å | $c = 16.281(2)$ Å | | | | | | |

| Atoms | Wyckoff | $x$ | $y$ | $z$ | $B_{\text{iso}}$ (Å$^2$) | Occ. | $M_z$ |
|---|---|---|---|---|---|---|---|
| Fe1 | 4e | 0.000 | 0.000 | 0.6720(3) | 0.23(3) | 1 | 1.6(2) |
| Fe2 | 2c | 0.667 | 0.333 | 0.750 | 0.4(1) | 0.86(2) | 1.6(2) |
| Ge | 2d | 0.333 | 0.667 | 0.750 | 1.73(9) | 1 | / |
| Te | 4f | 0.667 | 0.333 | 0.5891(3) | 0.24(4) | 1 | / |

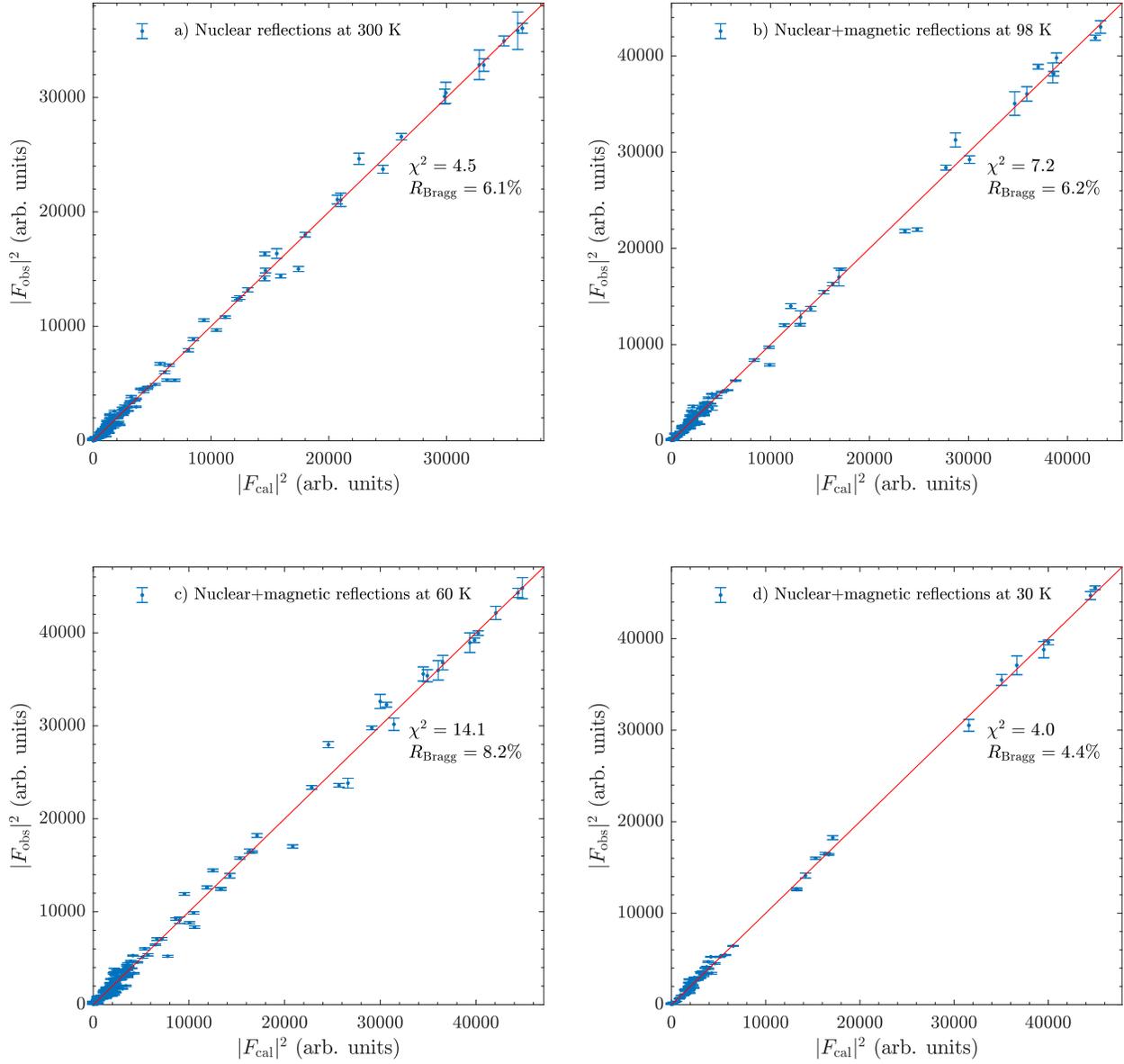

FIG. S1: Observed versus calculated intensities for: (a) nuclear reflections measured at 300 K; (b-d) both nuclear and magnetic reflections measured at (b) 98 K, (c) 60 K and (d) 30K.

changed. Figure S2(g) shows the difference image after the next 50mT field step. The magnetization of the sample has fully reversed compared to the image recorded at 150mT (Figure S2(b)).

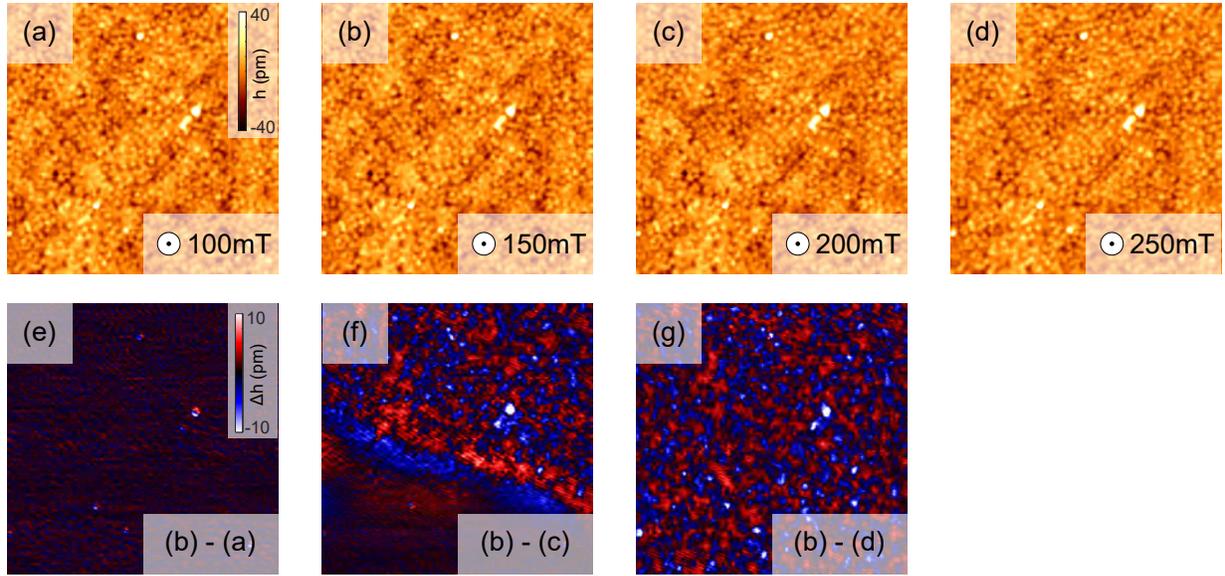

FIG. S2: (a) SP-STM topography (100mV, 40pA, $44 \times 44\text{nm}^2$) of the $Fe_3GeTe_2$ surface recorded in a field $B = 100\text{mT}$ along the crystal $c$ axis. (b) As (a) but recorded in a field $B = 150\text{mT}$. (c) As (a) but recorded in a field $B = 200\text{mT}$. A domain wall can be seen running diagonally through the image. (d) As (a) but recorded in a field $B = 250\text{mT}$. The magnetization of tip and sample spin are now opposite to those in image (a). (e) The difference of images (b) and (a). (f) The difference of images (b) and (c). The domain wall can clearly be seen to cross the image. (g) The difference between (b) and (d). The relative orientation of the magnetization of tip and sample has changed everywhere.

## S3. SPIN MODEL FOR $Fe_3GeTe_2$

In the main text, the data were modelled by a mapping to the simple hexagonal lattice nearest-neighbor Heisenberg model. The structure of $Fe_3GeTe_2$ is slightly more complicated than this, however there is a direct mapping between the two models in the limit of small $\mathbf{q}$. Moreover STM is sensitive to the surface, which forms an effective hexagonal lattice. We now perform the spin wave analysis for the more complicated case of the realistic $Fe_3GeTe_2$ lattice, showing that the small $\mathbf{q}$ expansion results in a difference of only a constant prefactor. The lattice is identical to the hexagonal lattice, except the B sites are duplicated, with one spin located at $+\delta z$ out of the plane and the other at $-\delta z$. We will call these additional sublattices $B_\text{u}$ and $B_\text{d}$ in the following. Consequently, the A and B sites now have a coordination of six and four respectively.

5

The spin wave Hamiltonian can now be written as

$$\mathcal{H} = J \sum_{\mathbf{r}}^{\substack{\alpha \in \{u,d\} \\ j \in \{0,1,2\}}} \mathbf{S}_A(\mathbf{r}) \cdot \mathbf{S}_{B_\alpha}(\mathbf{r} + \mathbf{a}_j) + J \sum_{\mathbf{r}} \mathbf{S}_{B_u}(\mathbf{r}) \cdot \mathbf{S}_{B_d}(\mathbf{r}) \\ + K \sum_{\mathbf{r}}^{\alpha \in \{u,d\}} [(\hat{S}_A^z(\mathbf{r}))^2 + (\hat{S}_{B_\alpha}^z(\mathbf{r}))^2]. \tag{S1}$$

we perform a Holstein-Primakoff transformation and so the Hamiltonian can be written in momentum space as

$$\mathcal{H} = \sum_{\mathbf{q}} \Big\{ \Big( \gamma_{\mathbf{q}} b_A^\dagger(\mathbf{q}) b_{B_u}(\mathbf{q}) + \gamma_{\mathbf{q}} b_A^\dagger(\mathbf{q}) b_{B_d}(\mathbf{q}) + JS b_{B_u}^\dagger(\mathbf{q}) b_{B_d}(\mathbf{q}) \Big) + \text{h.c.} \\ - 6JS b_A^\dagger(\mathbf{q}) b_A(\mathbf{q}) - 4JS \Big[ b_{B_u}^\dagger(\mathbf{q}) b_{B_u}(\mathbf{q}) + b_{B_d}^\dagger(\mathbf{q}) b_{B_d}(\mathbf{q}) \Big] \Big\} \\ - 2KS \sum_{\mathbf{q}} \Big\{ b_A^\dagger(\mathbf{q}) b_A(\mathbf{q}) + b_{B_u}^\dagger(\mathbf{q}) b_{B_u}(\mathbf{q}) + b_{B_d}^\dagger(\mathbf{q}) b_{B_d}(\mathbf{q}) \Big\}. \tag{S2}$$

As in the main text, we have defined $\gamma_{\mathbf{q}} = JS \sum_j e^{i\mathbf{q} \cdot \mathbf{a}_j}$. This Hamiltonian can be written in the compact form, $\mathcal{H} = \sum_{\mathbf{q}} \Psi_{\mathbf{q}}^\dagger \underline{\underline{M}} \Psi_{\mathbf{q}}$, where $\Psi_{\mathbf{q}}^\dagger = (b_A^\dagger(\mathbf{q}), b_{B_u}^\dagger(\mathbf{q}), b_{B_d}^\dagger(\mathbf{q}))$ and

$$\underline{\underline{M}} = \begin{pmatrix} -6JS & \gamma_{\mathbf{q}} & \gamma_{\mathbf{q}} \\ \gamma_{\mathbf{q}}^* & -6JS & JS \\ \gamma_{\mathbf{q}}^* & JS & -4JS \end{pmatrix}. \tag{S3}$$

This Hamiltonian can be diagonalized by a Bogoliubov transformation, or by diagonalizing the matrix $\mathbf{g}\underline{\underline{M}}$, where $\mathbf{g} = [\Psi_{\mathbf{q}}, \Psi_{\mathbf{q}}^\dagger]$ is the metric that enforces the bosonic commutation relations [2], which in our case is just the identity matrix

$$\mathbf{g} = [\Psi_{\mathbf{q}}, \Psi_{\mathbf{q}}^\dagger] = \begin{pmatrix} 1 & 0 & 0 \\ 0 & 1 & 0 \\ 0 & 0 & 1 \end{pmatrix}. \tag{S4}$$

After diagonalizing $\mathbf{g}\underline{\underline{M}}$, we find the dispersion relation for the two dispersive modes along $(H, H)$

$$\epsilon_{\pm}(H, H) = -\left(\frac{9}{2}JS + 2KS\right) \pm \frac{1}{2} JS \sqrt{33 + 32\cos(2\pi H) + 16\cos(4\pi H)}. \tag{S5}$$

Expanding the lower mode about $H = 0$, we find $\epsilon_+ \sim \frac{16\pi^2}{3} JSH^2$ at the bottom of the band, compared with $\epsilon_+ \sim 4\pi^2 JSH^2$ in the simple hexagonal model.

6

## S4. DERIVATION OF DOMAIN WALL PROFILE

We now present a derivation of the domain wall profile of a hexagonal lattice Heisenberg model. We will then show that the more realistic model of Fe$_3$GeTe$_2$ presented in Sect. S3 gives consistent results, justifying the use of the simple hexagonal model used in the main text.

### A. Continuum limit of spin model

The Heisenberg model on the hexagonal lattice, with nearest neighbor coupling can be written as

$$\mathcal{H} = J \sum_{\mathbf{r}\text{ (unit cells)}}^{j \in \{0,1,2\}} \mathbf{S}_A(\mathbf{r}) \cdot \mathbf{S}_B(\mathbf{r} + a\hat{\mathbf{e}}_j), \tag{S6}$$

where the sum is over all unit cells. By representing the spins as classical vectors of length $S$, we can develop a continuum description of the magnetization. We introduce the transformation $\mathbf{S}(\mathbf{r}) = S\mathbf{n}(\mathbf{r})$, dropping the site index in the spirit of a classical treatment of the spin degree of freedom. Neglecting constant terms, Eqn. S6 can be written as

$$\mathcal{H} = -\frac{JS^2}{2} \sum_{\mathbf{r}\text{ (unit cells)}}^{j \in \{0,1,2\}} \left[\mathbf{n}(\mathbf{r} + d_{\text{nn}}\hat{\mathbf{e}}_j) - \mathbf{n}(\mathbf{r})\right]^2. \tag{S7}$$

The basis vectors $\hat{\mathbf{e}}_j$ can readily be converted from the non-orthogonal $P6_3/mmc$ rhombohedral coordinate system into Cartesian coordinates. In Cartesian coordinates we have $\hat{\mathbf{e}}_1 = (-\frac{\sqrt{3}}{2}, \frac{1}{2})$, $\hat{\mathbf{e}}_2 = (\frac{\sqrt{3}}{2}, \frac{1}{2})$ and $\hat{\mathbf{e}}_3 = (0, -1)$ (Fig. S3), and hence

$$[\mathbf{n}(\mathbf{r} + d_{\text{nn}}\hat{\mathbf{e}}_1) - \mathbf{n}(\mathbf{r})] = -\frac{\sqrt{3}}{2} d_{\text{nn}} \partial_x(\mathbf{n}) + \frac{1}{2} d_{\text{nn}} \partial_y(\mathbf{n}) \tag{S8a}$$

$$[\mathbf{n}(\mathbf{r} + d_{\text{nn}}\hat{\mathbf{e}}_2) - \mathbf{n}(\mathbf{r})] = \frac{\sqrt{3}}{2} d_{\text{nn}} \partial_x(\mathbf{n}) + \frac{1}{2} d_{\text{nn}} \partial_y(\mathbf{n}) \tag{S8b}$$

$$[\mathbf{n}(\mathbf{r} + d_{\text{nn}}\hat{\mathbf{e}}_3) - \mathbf{n}(\mathbf{r})] = d_{\text{nn}} \partial_y(\mathbf{n}). \tag{S8c}$$

Performing the sum over $j$ we find

$$\mathcal{H} = -\frac{3JS^2}{2} \sum_{\mathbf{r}\text{ (unit cells)}} d_{\text{nn}} \left[(\partial_x \mathbf{n})^2 + (\partial_y \mathbf{n})^2\right]. \tag{S9}$$

Finally we take the continuum limit $\sum_{\mathbf{r}} \to \frac{1}{A_{unit}} \int d^2r$ to find a classical field theory of the Heisenberg model on the hexagonal lattice

$$\mathcal{H} = \frac{\rho_S}{2} \int d^2r \left\{(\partial_x \mathbf{n})^2 + (\partial_y \mathbf{n})^2\right\} \tag{S10}$$

7

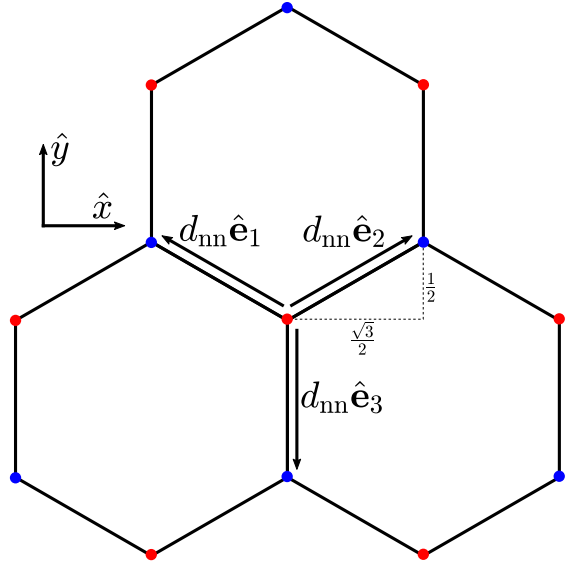

FIG. S3: Definition of the basis vectors $\hat{e}_i$ ($i = 1 \ldots 3$) for a honeycomb lattice.

where we have defined the spin stiffness $\rho_S = \frac{3JS^2 d_{nn}^2}{A_{unit}} = \frac{3JS^2 d_{nn}^2}{2A_{site}}$. Now let us perform the continuum limit of the single ion anisotropy term

$$\mathcal{H}_{\text{anis}} = K \sum_{\mathbf{r}}^{(\text{unit cells})} \left[ (S_A^z(\mathbf{r}))^2 + (S_B^z(\mathbf{r}))^2 \right] = KS^2 \sum_{\mathbf{r}}^{(\text{sites})} [n^z(\mathbf{r})]^2 \quad (S11)$$
$$= \kappa \int d^2r \left\{ (n^z(\mathbf{r}))^2 \right\}$$

with the anisotropy parameter $\kappa = \frac{2KS^2}{A_{\text{unit}}} = \frac{KS^2}{A_{\text{site}}}$.

### B. Domain wall solutions

The terms originating from exchange (Eqn. S10) can be seen as representing the stiffness of the spin field, penalizing the canting of spins away from ferromagnetic alignment. In a domain wall, these terms compete against the single-ion anisotropy terms (Eqn. S11), the former favoring a spatially extended domain wall such that the spins remain closely aligned, and the latter favoring a rotation of spins over a short length scale when $\theta$ is far from $0$ or $\pi$.

Following the general approach outlined in Ref. [3], we will now seek to derive an expression for the domain wall shape on the hexagonal lattice. We are interested in an one-dimensional cut through the domain wall, which we take to take to occur along $\hat{y}$ (Fig. S4), assuming that the spins rotate in a common rotation plane $(x, z)$ defined by the angle $\theta$, $\mathbf{n}(\mathbf{r}) = (\sin\theta(\mathbf{r}), \cos\theta(\mathbf{r}))$. In

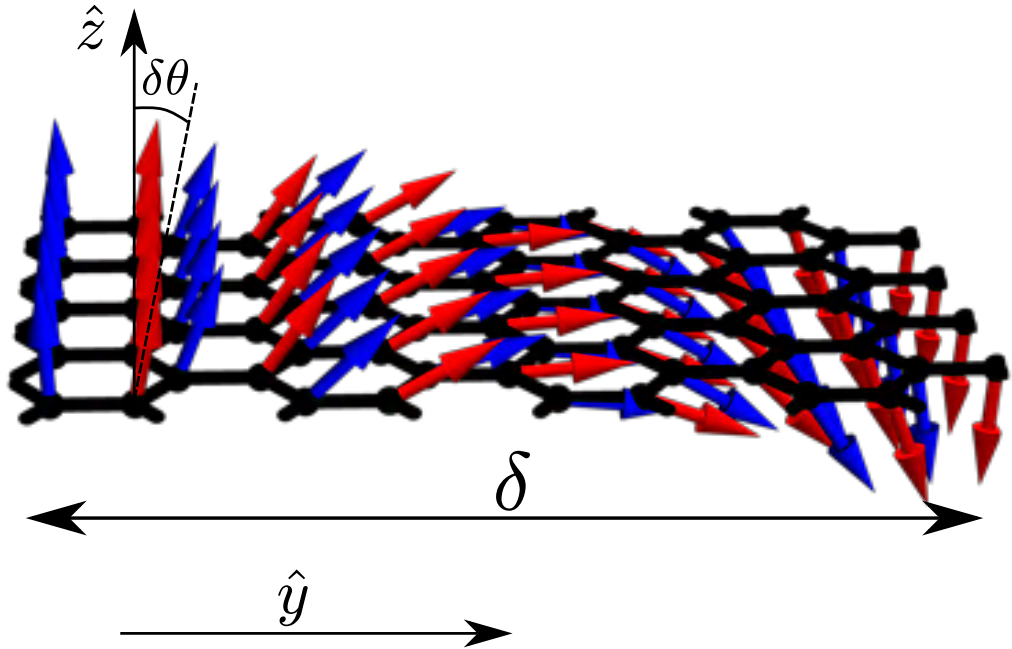

FIG. S4: Visualization of a Néel domain wall on the 2-d hexagonal lattice. The blue and red arrows indicate spins on the A and B sublattices respectively. The angle between the direction in which the spin is pointing locally and $\hat{z}$ is given by $\theta$. The canting angle between neighbouring spins is $\delta\theta$, which is smaller closer to the edges of the domain wall. Both $\theta$ and $\delta\theta$ are shown here for the second row of atoms parallel to the domain wall, where $\theta = \delta\theta$ (in the first row $\theta = 0$, and the spins are parallel to $\hat{z}$). Colors denote the sublattice. We note that the calculation for a Bloch-type domain wall results in the same expression for the domain wall width.

terms of this rotation angle, the energy of a domain wall can be written as

$$E_{DW} = \int d^2r \left\{ \frac{\rho_S}{2}(\partial_y \theta)^2 + \kappa \left(1 - \cos^2\theta\right) \right\}. \tag{S12}$$

Note here that we have taken $\partial_x \theta = 0$ as required for a domain wall along $\hat{y}$. Using the domain wall Lagrangian $\mathcal{L} = \frac{\rho_S}{2}(\partial_y \theta)^2 + \kappa \sin^2\theta$, the static Euler-Lagrange equation, $\frac{d}{dy}\left(\frac{\partial \mathcal{L}}{\partial \dot{\theta}}\right) = \frac{d\mathcal{L}}{d\theta}$, (where $\dot{\theta} = \frac{\partial \theta}{\partial y}$) can be written down and solved for the domain wall profile. Using the boundary condition $\frac{d\theta}{dy} \to 0$ as $y \to \pm\infty$, we find

$$\theta(y) = 2\arctan\left(e^{\sqrt{\frac{2\kappa}{\rho_S}}(y-y_0)}\right). \tag{S13}$$

This is known as the kink soliton, and reflects the conflict between the exchange stiffness energy

9

and the anisotropy energy in the domain wall. The spins rotate little between nearest neighbours at the edges of the domain wall, where the anisotropy energy is small, before rotating much more between adjacent sites when this energy increases at the center of the wall. After some algebra, the $z$-component of the spin field can be written down

$$n^z(y) = \tanh\left[\sqrt{\frac{2\kappa}{\rho_S}}(y-y_0)\right]$$
$$= \tanh\left[\sqrt{\frac{4K}{3Jd_{\mathrm{nn}}^2}}(y-y_0)\right] \tag{S14}$$

which takes the shape of a step function with a width that depends on $\sqrt{\frac{J}{K}}$. We can rewrite this expression as

$$n^z(y) = \tanh(\pi y/\delta) \tag{S15}$$

for a domain wall at $y_0 = 0$, defining its intrinsic width $\delta = \pi\sqrt{\frac{\rho_S}{2\kappa}}$ which is the same as the one obtained under the assumption of a linear variation of $\theta(y)$ [4] and as used in previous work [5].

### C. Equivalence of hexagonal and Fe$_3$GeTe$_2$ model

Finally we will show the equivalence of modelling the data using both the hexagonal model and the more realistic model presented in S3 so long as the choice of model is consistent for both the neutron and the STM data.

The spin model for the full crystallographic structure of Fe$_3$GeTe$_2$ is given by Eqn. S1. As in the previous section, we represent the spins in the continuum limit by a vector field $\mathbf{S}(\mathbf{r}) = S\mathbf{n}(\mathbf{r})$, dropping the site index and assuming that the spins on the $u$ and $d$ sublattices are coaligned. Neglecting constant terms we have

$$\mathcal{H} = -JS^2 \sum_{\mathbf{r} \text{ (unit cells)}}^{j \in \{0,1,2\}} [\mathbf{n}(\mathbf{r}+d_{\mathrm{nn}}\hat{\mathbf{e}}_j) - \mathbf{n}(\mathbf{r})]^2 \tag{S16a}$$

$$\mathcal{H}_{anis} = \frac{3}{2}KS^2 \sum_{\mathbf{r}}^{\text{(sites)}'} [n^z(\mathbf{r})]^2 \tag{S16b}$$

where $'$ indicates that the sum is performed over the sites of the hexagonal lattice since the sum over the $u$ and $d$ sublattices has already been explicitly performed. By comparing these expressions with the analogous expressions for the hexagonal model Eqns. S7 and S11, it is possible to map

from one model to the other,

$$\rho_S^{hex} = \frac{1}{2}\rho_S^{FGT} \tag{S17a}$$

$$\kappa^{hex} = \frac{2}{3}\kappa^{FGT}. \tag{S17b}$$

thus $\delta^{FGT} = \sqrt{\frac{4}{3}}\delta^{hex}$. Examination of the small-**q** expansion of the dispersion along (H,H) for the two models

$$\epsilon_+^{hex} \approx -2KS - 4\pi^2 J_{hex} SH^2 \tag{S18a}$$

$$\epsilon_+^{FGT} \approx -2KS - \frac{16}{3}\pi^2 J_{FGT} SH^2 \tag{S18b}$$

shows that $J_{hex} = \frac{4}{3}J_{FGT}$, therefore any consistent treatment of the STM and neutron data using either model, as described above, yields consistent results.

## S5. ADDITIONAL DENSITY FUNCTIONAL THEORY CALCULATIONS

In Figure S5 we show the orbitally projected band structure obtained from the DFT calculations. It can be seen that the bands of majority spin character have predominantly Te character, where the ones of minority spin character have more weight on the Fe atoms. Since the surface atoms are Te atoms, this suggests that the STM will predominantly couple to bands of majority spin character.

[1] S. Chris, A. K. Nebil, C. Edmond, F. R. O. Ramon, L. Harry, P. Elise, and Q. Navid, *The nuclear structure of CVT grown Fe3-xGeTe2*, type: dataset, URL https://doi.ill.fr/10.5291/ILL-DATA.5-11-440.

[2] R. M. White, M. Sparks, and I. Ortenburger, Phys. Rev. **139**, A450 (1965), URL https://link.aps.org/doi/10.1103/PhysRev.139.A450.

[3] A. Aharoni, *Introduction to the Theory of Ferromagnetism*, International Series of Monographs on Physics (Clarendon Press, 2000), ISBN 9780198508090, URL https://books.google.co.uk/books?id=Ru-z9b3WcfMC.

[4] S. Blundell, *Magnetism in condensed matter* (Oxford ; New York : Oxford University Press, 2001., 2001), includes bibliographical references and index., URL https://search.library.wisc.edu/catalog/999921680802121.

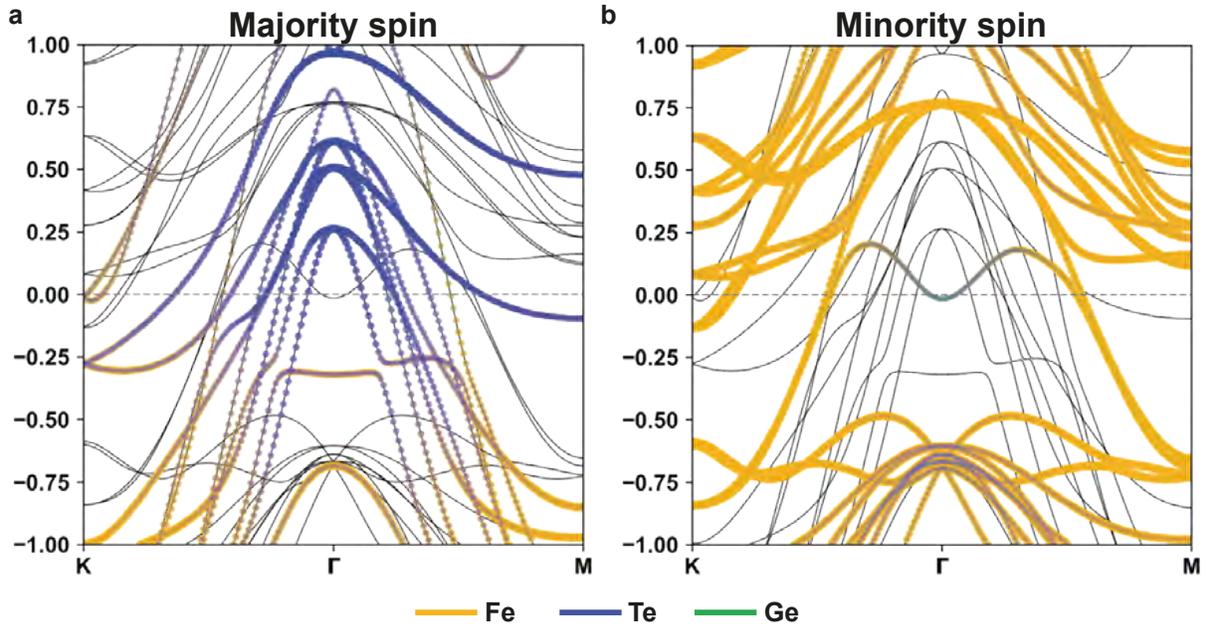

FIG. S5: Orbital projections of bulk Fe$_3$GeTe$_2$ from density functional theory calculations. a) Orbital projection of the majority bands. b) Orbital projection of the minority bands. the majority spin is dominated by Te weight, whereas the minority spin is dominated by Fe weight. There is very little Ge weight situated around the Fermi level.

[5] R. Moreno, R. F. L. Evans, S. Khmelevskyi, M. C. Muñoz, R. W. Chantrell, and O. Chubykalo-Fesenko, Phys. Rev. B **94**, 104433 (2016), URL https://link.aps.org/doi/10.1103/PhysRevB.94.104433.


\end{document}